\documentclass[aps,prd,groupedaddress,showpacs,nofootinbib,amssymb,balancelastpage,preprintnumbers,floatfix]{revtex4}

\usepackage[dvipdfmx]{graphicx}
\usepackage{graphicx,bm,color}

\usepackage{amsmath}
\usepackage{amssymb}
\usepackage{amsfonts}
\usepackage{cases}
\usepackage{inconsolata}
\usepackage{braket}
\usepackage{array}
\usepackage{booktabs} 

\usepackage{cancel}
\usepackage{hyperref}
\usepackage[normalem]{ulem}
\usepackage{subfigure}
\usepackage{here}
\usepackage{color}
\usepackage{comment}
\usepackage[english]{babel}
\usepackage{slashed}
\usepackage{tikz}
\usetikzlibrary{matrix}

\allowdisplaybreaks[3]
\usepackage{subfigure} 
\allowdisplaybreaks[3]

\begin{document}

\title{Ladder top-quark condensation imprints in supercooled electroweak phase transition} 

\author{Yuepeng Guan}\thanks{{\tt guanyp22@mails.jlu.edu.cn}}
\affiliation{Center for Theoretical Physics and College of Physics, Jilin University, Changchun, 130012,
	China}

\author{Shinya Matsuzaki}\thanks{{\tt synya@jlu.edu.cn}}
\affiliation{Center for Theoretical Physics and College of Physics, Jilin University, Changchun, 130012,
	China}%

\begin{abstract}
The electroweak (EW) phase transition in the early Universe might be supercooled due to the presence of the classical scale invariance involving Beyond the Standard Model (BSM) sectors and the supercooling could persist down till a later epoch around which the QCD chiral phase transition is supposed to take place. Since this supercooling period keeps masslessness for all the six SM quarks, it has simply been argued that the QCD phase transition is the first order, and so is the EW one. However, not only the QCD coupling but also the top Yukawa and the Higgs quartic couplings get strong at around the QCD scale due to the renormalization group running, hence this scenario is potentially subject to a rigorous nonperturbative analysis. In this work, we employ the ladder Schwinger-Dyson (LSD) analysis based on the Cornwall-Jackiw-Tomboulis formalism at the two-loop level in such a gauge-Higgs-Yukawa system.  
We show that the chiral broken QCD vacuum emerges with the nonperturbative top condensate and the lightness of all six quarks is guaranteed due to the accidental U(1) axial symmetry presented in the top-Higgs sector. We employ a quark-meson model-like description in the mean field approximation to address the impact on the EW phase transition arising due to the top quark condensation at the QCD phase transition epoch. In the model, the LSD results are encoded to constrain the model parameter space. 
We then observe the cosmological phase transition of the first-order type 
and discuss the induced gravitational wave (GW) productions. We find that in addition to 
the conventional GW signals sourced from an expected BSM at around or over the TeV scale, the dynamical topponium-Higgs system can yield another power spectrum sensitive to the BBO, LISA, and DECIGO, etc. 

\end{abstract}

\maketitle

\section{Introduction}

The QCD phase transition is of particular importance to understand the origin of mass and property for matter, i.e., nucleon. It has extensively been explored so far based on lattice simulations as well as low-energy effective field theories. However, in application to cosmology, the QCD phase transition in the Hubble expanding universe cannot be observed in lattice QCD or experiments today. Therefore, actually, we have little understood the QCD cosmology in the sense of the thermal history of the universe.

%Deep investigation of the QCD cosmology thus expects to shed a %new light also on what the Beyond the Standard Model (BSM) would %be like. Collider experimental sensitivities to detect the BSM %related to the low-energy QCD physics is lower than the other %sectors in the SM, because of huge hadronic background events %created until reaching the detectors. This also implies the new %physics window as a currently unexplored ``loophole" in low-%energy QCD physics. This might hide existence of dark matters %and/or some evidence of the dark side of the universe related to %the matter-anti-matter asymmetry.  

  The recent and prospected observations of nano hertz gravitational waves (GWs) have the potential sensitivity to deeply clarify the thermal history around the QCD phase transition epoch. The observation of a stochastic GW background has been reported from the NANOGrav pulsar timing array (PTA) collaboration in 15 years of data~\cite{NANOGrav:2023gor,NANOGrav:2023hfp}. Possible origins of the detected nano Hz peak frequency in a view of the BSM have been investigated also by the NANOGrav collaboration \cite{NANOGrav:2023hvm}. Other recent PTA data, such as those from the European PTA (EPTA)\cite{EPTA:2023sfo,EPTA:2023akd,EPTA:2023fyk}, Parkes PTA (PPTA)\cite{Reardon:2023gzh,Reardon:2023zen}, and Chinese PTA (CPTA)~\cite{Xu:2023wog} have also supported the presence of consistent nano-hertz stochastic GWs. 
  The tail of the GW spectra produced at the QCD phase transition epoch still keeps having a high enough sensitivity also at other interferometers designated aiming at the higher frequency spectra, like the Laser Interferometer Space Antenna (LISA)~\cite{LISA:2017pwj,Caprini:2019egz}, 
the Big Bang Observer (BBO)~\cite{Corbin:2005ny,Harry:2006fi}, and Deci-hertz Interferometer Gravitational Wave Observatory (DECIGO)~\cite{Kawamura:2006up,Yagi:2011wg}, etc. 
Thus, such GW evidence is anticipated to provide us with a clue on the new aspect of the QCD cosmology in the thermal history of the universe.

The QCD phase transition has been confirmed to be crossover by  
the lattice QCD with 2 + 1 flavors at the physical point 
at the pseudocritical temperature $T_{\rm pc}\sim 155$ MeV~\cite{Aoki:2009sc,Borsanyi:2011bn,Ding:2015ona,HotQCD:2018pds,Ding:2020rtq} for the chiral phase transition,  
and at the same time the deconfinement-confinement transition is expected 
to take place as well~\cite{Bazavov:2016uvm,Ding:2017giu}.  
This is the current consensus of the QCD phase transition to our best knowledge.  
In the Hubble evolutionary 
universe with BSM, however, the cosmological QCD phase transition 
might be more involved and richer.

One particular interest is in when the SM is extended to be 
scale-invariant along with a dark sector, 
subsequent supercooling including the QCD phase transition 
can be realized~\cite{Iso:2017uuu,Hambye:2018qjv,vonHarling:2017yew}.
In this class of the scenario context, the QCD phase transition also plays an important role in triggering the electroweak (EW) phase transition, which can be strong first-order at temperatures  
$T = {\cal O}( T _{\rm pc})$, thus the EW phase transition can be 
supercooled until then. 
In the literature  
several phenomenological and cosmological consequences arising from this non-standard QCD-scale cosmology have been discussed including the impact on the  baryogenesis~\cite{Dichtl:2023xqd,Ellis:2022lft,Wang:2022ygk}, formation of compact star objects~\cite{DelGrosso:2024wmy}, and the gravitational wave detection sensitivities mainly sourced from the scale-invariant dark sector along with or without a dark matter production~\cite{Iso:2017uuu,Hambye:2018qjv,Sagunski:2023ynd,Dichtl:2023xqd,Frandsen:2022klh,Ellis:2022lft,Bodeker:2021mcj,vonHarling:2017yew,Wong:2023qon}.

\begin{figure}[H]
\centering
\subfigure[]{
\label{(a)}
\includegraphics[width=0.45\textwidth]{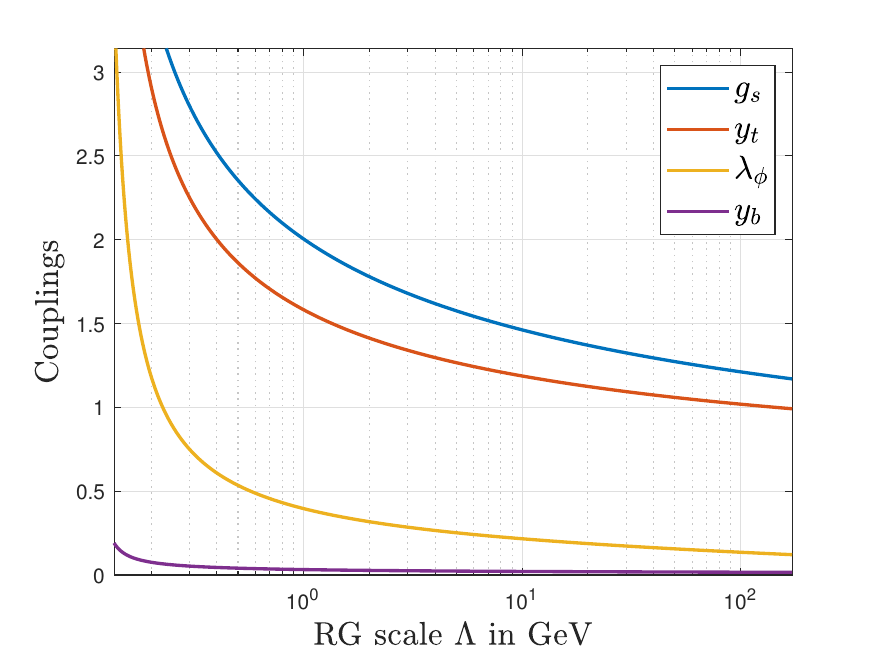}}
\quad
\subfigure[]{
\label{(b)}
\includegraphics[width=0.45\textwidth]{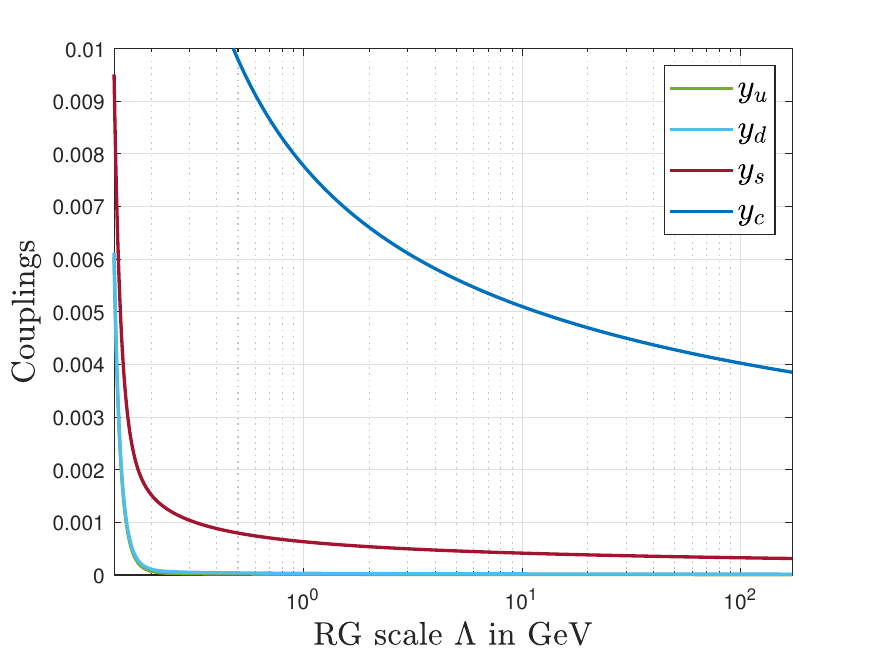}}
\caption{The perturbative two-loop running in the classically scale-invariant SM for the Yukawa couplings (denoted as $y_f$ with $f$ being fermion flavor label), the Higgs quartic coupling ($\lambda_\phi$), and 
the QCD gauge coupling $g_s$. 
Here inputs have been made of $m_h = 125.25$ GeV 
(the Higgs mass);  
$v_\text{EW} = 246$ GeV (the EW scale); 
$m_Z = 91.1876$ GeV (the Z boson mass); 
$\alpha_s(m_Z) (\equiv g_s^2/(4\pi)) = 0.1184$ 
(the $\overline{\text{MS}}$ QCD coupling renormalized at the $m_Z$ scale);  $m_t = 172.5$ GeV (the top quark mass); 
$m_b = 4.18$ GeV and  
$m_c = 1.27$ GeV (the $\overline{\text{MS}}$ masses renormalized at their $\overline{\text{MS}}$ masses); 
$m_s = 93.4 \times 10^{-3}$ GeV, $m_d = 4.67 \times 10^{-3}$ GeV, 
and $m_u = 2.16 \times 10^{-3}$ GeV (the $\overline{\text{MS}}$ masses renormalized at 2 GeV). All the inputs above follow the Particle Data Group~\cite{ParticleDataGroup:2022pth}. 
%(a)the bottom-Yukawa coupling $y_b$ (in green), the QCD gauge coupling %$g_s$ (blue), the Higgs quartic coupling $\lambda_\phi$ (red), and top %Yukawa coupling $y_t$ (blue line); (b)Up Yukawa $y_u$ (blue line), down %Yukawa $y_d$ (red line), strange Yukawa $y_s$ (yellow line) and charm %Yukawa $y_c$ (purple line).
}
\label{RG-SM}
\end{figure}

However, the quantum field theory in the thus supercooled epoch may not merely be governed by QCD: the top Yukawa and the Higgs quartic couplings as well as the QCD gauge coupling get strong at around the QCD scale due to the renormalization group running, 
as plotted in Fig.~\ref{RG-SM}. 
Hence this scenario is potentially subject to a rigorous nonperturbative analysis, which has never been addressed at this point.

 In this paper, we make the first attempt to employ the nonperturbative analysis of the QCD-induced EW phase transition scenario. 
The method that we apply is the ladder Schwinger-Dyson (LSD) equation based on the Cornwall-Jackiw-Tomboulis (CJT) formalism~\cite{Cornwall:1974vz} at the two-loop level.  
Working on the gauge-Higgs-Yukawa system including QCD, the Higgs-Yukawa, and the Higgs quartic interactions, 
we show that the chiral broken QCD vacuum emerges with the nonperturbative top condensate and the lightness of all six quarks is guaranteed. 
The latter turns out to be due to the accidental U(1) axial symmetry 
(approximately) present in the top-Higgs Yukawa sector (with small enough EW gauge interactions neglected).

We discuss the impact on the EW phase transition in the thermal history arising from the emergence of the top quark condensation in the QCD phase transition epoch. 
We monitor the low-energy effective theory by a quark-meson model-like description, for the top - Higgs hybrid system. 
The nonperturbative scale anomaly induced from the dynamical top mass generation based on the LSD analysis is encoded in the model by the anomaly matching procedure, to constrain the model parameter space. 
We then observe the cosmological phase transition of the first-order type 
 in the mean field approximation and discuss the associated GW productions. 
 We find that in addition to 
the conventional GW signals sourced from an expected BSM at around or over the TeV scale, the dynamical topponium-Higgs system can yield another power spectrum sensitive to the BBO, LISA, and DECIGO, etc.

This paper is structured as follows. 
In Sec.~II we start with deriving the SD equations in the gauge-Higgs-Yukawa system relevant to the presently concerned supercooling scalegenesis, and introduce the ladder approximation. In Sec.~III,  the LSD equations are reproduced in the framework of the CJT formalism at the two-loop level, which also provides the nonperturbative scale anomaly induced from the nontrivial solutions to the LSD equations including the top and Higgs condensates. 
Then the phase structure with the scale-invariant SM prediction is discussed in detail. 
In Sec.~IV, we model a low-energy description of the present 
gauge-Higgs-Yukawa theory to be a quark-meson model-like, to discuss the top quark condensation on the effect on the supercooled EW phase in the thermal history. Employing the mean-field approximation, we evaluate the associated cosmological phase transition in the top - Higgs hybrid system and explore the GW production and signals. The summary of the present paper is presented in Section V, along with discussions related to issues to be pursed in the future.

\section{The SD formalism in the gauge - Higgs - Yukawa system: gauge fixing in the ladder approximation} 

In this section, we derive the exact SD equations in the currently 
concerned gauge - Higgs - Yukawa system and discuss the ladder truncation. 
The gauge-Higgs Yukawa system, which 
dominates the low-energy description of the classically scale-invariant SM, is described by the following Lagrangian: 
\begin{equation}
    \mathcal{L} = \mathcal{L}_\text{gauge} + \mathcal{L}_\text{quarks} + \mathcal{L}_\text{Higgs} + \mathcal{L}_\text{Yukawa},
\label{GHY-Lag}
\end{equation}
where
\begin{equation}
    \begin{aligned}
        %&\mathcal{L}_\text{gauge} = -\frac{1}{4}F^a_{\mu\nu}F^{a,\mu\nu} 
        %+ \mathcal{L}_\text{guage fix} + \mathcal{L}_\text{ghost},\\
        &\mathcal{L}_\text{quark} = i \bar{q_i}\left( \slashed{\partial} + ig_s\slashed{A} \right) q_i,\\
        &\mathcal{L}_\text{Higgs} = (\partial_\mu H)^\dagger \partial^\mu H  - \lambda_\phi( H^\dagger H )^2,\\
        &\mathcal{L}_\text{Yukawa} = -y_t \bar{q_3}_L \Tilde{H} t_R -y_b \bar{q_3}_L H b_R +\text{h.c.}
        \,, 
    \end{aligned}
\end{equation}
and ${\cal L}_{\rm gauge}$ includes the QCD gauge kinetic and 
Lorentz-covariant gauge fixing terms ($- \frac{1}{2\xi} (\partial_\mu A^a_\mu)^2$) with the gluon fields $A_\mu^a$ ($a=1,\cdots, 8$) and the gauge fixing parameter $\xi$.  
In Eq.(\ref{GHY-Lag}) ${q_i}_L$ $(i=1,2,3)$ denotes the $i$th generation-$SU(2)_L$ quark doublet;  
%\begin{equation}
%    {q_i}_L = {q_i}_L^\alpha, \quad\quad i = 1, 2, 3 \quad %\text{and} \quad \alpha = 1,\cdots,N_c,
%\end{equation} 
$H$ denotes the Higgs doublet parameterized as 
%\begin{equation} 
$    H = 
    \left(
    \begin{array}{cc}
        \phi^+\\
        \phi^0
    \end{array}
    \right)
    =
    \frac{1}{\sqrt{2}}\left(
    \begin{array}{cc}
        \pi_2-i\pi_1 \\
        \sigma +i\pi_3
    \end{array}
    \right)
    $, and its charge conjugate field $\tilde{H}$ is defined as $\tilde{H} \equiv i \tau_2 H^*$ with $\tau_2$ being the Pauli matrix.  
    The Higgs (vacuum expectation value) VEV $\sigma_h$ arises as 
    $\sigma = \sigma_h + \widetilde{\sigma}$, with 
    $\sigma_h \equiv \sqrt{2 \braket{H^\dag H}} $. 
%\end{equation}
Only the third-generation quarks (top and bottom) 
become strongly coupled to the Higgs as well as to QCD, as seen from the renormalization group evolution in Fig.~\ref{RG-SM}, while 
other lighter quarks strongly couple only to QCD. 
The present analysis extends the earlier work in~\cite{Kondo:1993ty} in the sense that 
the isospin breaking in the Yukawa sector will be taken into account.

\subsection{The exact SD equations}

\begin{figure}[t]
  \centering
  \begin{tabular}{ccccccc}
  $iS^{-1}$ &
  =&
  $i{S_0}^{-1}$ &
  {\large +}&
  \includegraphics[width=40mm]{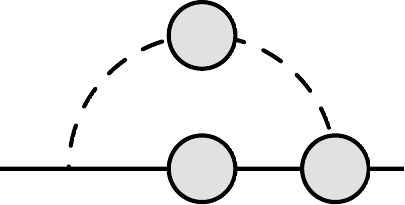} &
  {\large +}&
  \includegraphics[width=40mm]{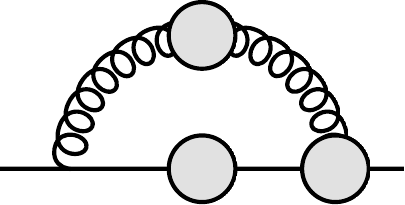}\\[6ex]
  \end{tabular}
  \caption{Feynman diagrams of the exact SD equations for the $t$ and $b$ quark propagators corresponding to Eqs.(\ref{SD-t}) and (\ref{SD-b}). $S_0^{-1}$ in the first term of the right-hand side stands for the tree-level inverse propagators, $i S_0^{-1}(p)|_{t,b} = \slashed{p} - \frac{y_{t,b}}{\sqrt{2}} \sigma_h$.  The blobs include all possible quantum corrections in the present gauge-Higgs-Yukawa system.} 
\label{exact-SD}
\end{figure}

From the Lagrangian~\ref{GHY-Lag} the exact SD equations for the inverse-full propagators for $t$ and $b$ quark fields, $S_t^{-1}$ and $S_b^{-1}$, 
are read off as (for Feynman graphical interpretations, see also Fig.~\ref{exact-SD}) 
\begin{align} 
        i S_\text{t}^{-1} &= \slashed{p} - \frac{y_t}{\sqrt{2}} \sigma_h - i \int \frac{d^4q}{(2\pi)^4}\, \frac{-iy_t}{\sqrt{2}}S_\text{t}(q) \Gamma^{(3)}_{\bar{t} \sigma t}(p,q) \Delta_{\sigma}(p-q) \notag\\
         &\hspace{20pt}- i \int \frac{d^4q}{(2\pi)^4}\, \frac{-\gamma_5 y_t}{\sqrt{2}}S_\text{t}(q) \Gamma^{(3)}_{\bar{t} \pi_3 t}(p,q) \Delta_{\pi_3}(p-q) \notag \\
         &\hspace{20pt}- i \int \frac{d^4q}{(2\pi)^4}\, \frac{- (y_b P_R - y_t P_L)}{\sqrt{2}}S_\text{b}(q) \left[ \Gamma^{(3)}_{\bar{b} \pi_1 t,R}(p,q)+\Gamma^{(3)}_{\bar{b} \pi_1 t,L}(p,q) \right] \Delta_{\pi_1}(p-q) \notag \\
         &\hspace{20pt}- i \int \frac{d^4q}{(2\pi)^4}\, \frac{-i 
         (y_b P_R - y_t P_L)}{\sqrt{2}}S_\text{b}(q) 
         \left[ \Gamma^{(3)}_{\bar{b} \pi_2 t,R}(p,q)+\Gamma^{(3)}_{\bar{b} \pi_2 t,L}(p,q) \right] \Delta_{\pi_2}(p-q) \notag \\
         &\hspace{20pt}- i \int \frac{d^4q}{(2\pi)^4}\, (-i g_s \gamma^\mu T^a)S_\text{t}(q) \Gamma^{(3),\nu}_{\bar{t} A t,a}(p,q) \Delta_{g,\mu\nu}(p-q)\,, 
    \label{SD-t:0}
\end{align} 
and 
    \begin{align}
        i S_\text{b}^{-1} &= \slashed{p} - \frac{y_b}{\sqrt{2}} \sigma_h - i \int \frac{d^4q}{(2\pi)^4}\, \frac{-iy_b}{\sqrt{2}}S_\text{b}(q) \Gamma^{(3)}_{\bar{b} \sigma b}(p,q) \Delta_{\sigma}(p-q) \notag \\
         &\hspace{20pt}- i \int \frac{d^4q}{(2\pi)^4}\, \frac{\gamma_5 y_b}{\sqrt{2}}S_\text{b}(q) \Gamma^{(3)}_{\bar{b} \pi_3 b}(p,q) \Delta_{\pi_3}(p-q) \notag \\
         &\hspace{20pt}- i \int \frac{d^4q}{(2\pi)^4}\, \frac{-(y_t P_R - y_b P_L)}{\sqrt{2}}S_\text{t}(q) \left[ \Gamma^{(3)}_{\bar{t} \pi_1 b,R}(p,q)+\Gamma^{(3)}_{\bar{t} \pi_1 b,L}(p,q) \right] \Delta_{\pi_1}(p-q) \notag \\
         &\hspace{20pt}- i \int \frac{d^4q}{(2\pi)^4}\, \frac{i( y_t P_R - y_b P_L)}{\sqrt{2}}S_\text{t}(q) \left[ \Gamma^{(3)}_{\bar{t} \pi_2 b,R}(p,q)+\Gamma^{(3)}_{\bar{t} \pi_2 b,L}(p,q) \right] \Delta_{\pi_2}(p-q) \notag \\
         &\hspace{20pt}- i \int \frac{d^4q}{(2\pi)^4}\, (-i g_s \gamma^\mu T_a)S_\text{b}(q) \Gamma^{(3),\nu}_{\bar{b} A b,b}(p,q) \Delta_{g,\mu\nu}^{ab}(p-q)\, , 
\label{SD-b:0}
    \end{align}
where $\Gamma^{(3)}_{\bar{\psi_i} \phi \psi_j}(p,q)$ represent the three-point vertex functions in the momentum space among the fermion field pair $\bar{\psi_i}$, $\psi_j$, and the boson field $\phi$; $T_a$ accompanied with the QCD vertex stands for the generator of $SU(3)$; 
the labels $L$ and $R$ denote the chirality projection involved in the vertex functions, defined like $\Gamma^{(3)}_{L/R} \equiv \Gamma^{(3)} P_{L/R}$ with $P_{L/R} \equiv (1 \mp \gamma_5)/2$;  
$\Delta_{\phi, g}$ denote the full propagators of boson fields. 
These equations with $y_t = y_b$ taken are smoothly reduced to the global chiral $U(2)_L \times U(2)_R$ invariant case which has been addressed in the literature~\cite{Kondo:1993ty}.

The SD equation for the Higgs VEV is evaluated at the leading order of 
the Yukawa interactions, which is dominated by the top-quark contribution, to be 
\begin{equation}
    \lambda_\phi \braket{\sigma(\sigma^2 + \vec{\pi}^2)} = 
    -\frac{y_t}{\sqrt{2}} \braket{\bar{t}t}\,, 
\label{SD-HiggsVEV}
\end{equation}
where $\vec{\pi} = (\pi_1, \pi_2, \pi_3)$.

\subsection{The ladder truncation and gauge fixing} 

We apply the ladder approximation for QCD, 
such that 
the gauge-fermion-fermion vertex functions are taken to be of tree level: 
\begin{equation}
        \Gamma^{(3),\nu}_{\bar{t} A t,a}(p,q) = \Gamma^{(3),\nu}_{\bar{b} A b,a}(p,q) = -i g_s \gamma^\nu T_a. 
\end{equation}
The abelian-type Ward-Takahashi identity with this truncation constrains the fermion full propagators as~\cite{Miransky:1994vk}  
\begin{equation}
        iS_{t/b}^{-1} = A_{t/b}(-p^2) \slashed{p} - B_{t/b}(-p^2)\,,  
        \qquad {\rm with} \qquad 
        A_{t/b}(-p^2) = 1 
        \,. \label{A1}
\end{equation}
It is well-known that in the pure QCD case, 
this condition can be fulfilled 
by taking the Landau gauge: $\xi=0$. 
The present case is, however, more involved, which also includes the Yukawa and Higgs quartic interactions.  
First of all, we therefore examine the consistency of the Landau gauge choice for the present system to satisfy Eq.(\ref{A1}). 
As it will turn out right below,  the extended ladder approximation 
with the bare propagators for both scalar and gauge fields 
in the Landau gauge 
is shown to be numerically consistent well with Eq.(\ref{A1}).

To prove this point, we take the tree-level propagators for the scalar and gauge fields as 
\begin{equation}
    \begin{aligned}
        &\Delta_\sigma(k) = \frac{i}{k^2 - 3\lambda_\phi \sigma_h^2},\\
        &\Delta_{\pi_i}(k) = \frac{i}{k^2 - \lambda_\phi \sigma_h^2},\\
        &\Delta_{g,\mu\nu}^{ab}(k) = \frac{-i \delta^{ab}}{k^2} \left( g_{\mu\nu} - (1-\xi)\frac{k_\mu k_\nu}{k^2} \right),\\
    \end{aligned}
\end{equation}
and all the relevant vertices to be bare ones:  
\begin{equation}
    \begin{aligned}
        &\Gamma^{(3)}_{\bar{t} \sigma t}(p,q) = \frac{-iy_t}{\sqrt{2}}, \quad \Gamma^{(3)}_{\bar{t} \pi_3 t}(p,q) = \frac{-\gamma_5 y_t}{\sqrt{2}},\\
        &\Gamma^{(3)}_{\bar{t} \pi_1 b,R}(p,q) = \frac{- y_b P_R}{\sqrt{2}}, \quad \Gamma^{(3)}_{\bar{t} \pi_2 b,R}(p,q) = \frac{-i y_b P_R}{\sqrt{2}},\\
        &\Gamma^{(3)}_{\bar{t} \pi_1 b,L}(p,q) = \frac{y_t P_L}{\sqrt{2}}, \quad \Gamma^{(3)}_{\bar{t} \pi_2 b,L}(p,q) = \frac{i y_t P_L}{\sqrt{2}},\\
    \end{aligned}
\end{equation}
\begin{equation}
    \begin{aligned}
        &\Gamma^{(3)}_{\bar{b} \sigma b}(p,q) = \frac{-iy_b}{\sqrt{2}}, \quad \Gamma^{(3)}_{\bar{b} \pi_3 b}(p,q) = \frac{\gamma_5 y_b}{\sqrt{2}},\\
        &\Gamma^{(3)}_{\bar{b} \pi_1 t,R}(p,q) = \frac{- y_t P_R}{\sqrt{2}}, \quad \Gamma^{(3)}_{\bar{b} \pi_2 t,R}(p,q) = \frac{i y_t P_R}{\sqrt{2}},\\
        &\Gamma^{(3)}_{\bar{b} \pi_1 t,L}(p,q) = \frac{y_b P_L}{\sqrt{2}}, \quad \Gamma^{(3)}_{\bar{b} \pi_2 t,L}(p,q) = \frac{-i y_b P_L}{\sqrt{2}}. \\
    \end{aligned}
\end{equation}
With this prescription, after a tedious computation, the $A_{t/b}(p_E^2)$ and $B_{t/b}(p_E^2)$ 
functions (with $p_E^2 \equiv - p^2$)  take the form 
\begin{equation}
    \begin{aligned}
        B_t(p_E^2) =& \frac{y_t}{\sqrt{2}} \sigma_h + \int^{\Lambda^2} d q_E^2 \mathcal{K}_{B_t}(p_E^2,q_E^2)\frac{q_E^2 B_t(q_E^2)}{ A_t^2(q_E^2)q_E^2 + B_t^2(q_E^2)}\\
        &+ \frac{y_t y_b}{16\pi^2}\int^{\Lambda^2} d q_E^2 K_B(p_E^2,q_E^2;\lambda_\phi \sigma_h^2)\frac{q_E^2 B_b(q_E^2)}{ A_b^2(q_E^2)q_E^2 + B_b^2(q_E^2) },\\
        A_t(p_E^2) =& 1 + \int^{\Lambda^2} d q_E^2 \mathcal{K}_{A_t}(p_E^2,q_E^2)\frac{A_t(q_E^2)}{ A_t^2(q_E^2)q_E^2 + B_t^2(q_E^2)}\\
        &+ \frac{{y_t}^2 + {y_b}^2}{32\pi^2}\int^{\Lambda^2} d q_E^2 K_A(p_E^2,q_E^2;\lambda_\phi \sigma_h^2)\frac{A_b(q_E^2)}{ A_b^2(q_E^2)q_E^2 + B_b^2(q_E^2) } ,
    \end{aligned}
\label{SD-t}
\end{equation}
\begin{equation}
    \begin{aligned}
        B_b(p_E^2) =& \frac{y_b}{\sqrt{2}} \sigma_h + \int^{\Lambda^2} d q_E^2 \mathcal{K}_{B_b}(p_E^2,q_E^2)\frac{q_E^2 B_t(q_E^2)}{ A_b^2(q_E^2)q_E^2 + B_b^2(q_E^2)}\\
        &+ \frac{y_t y_b}{16\pi^2}\int^{\Lambda^2} d q_E^2 K_B(p_E^2,q_E^2;\lambda_\phi \sigma_h^2)\frac{q_E^2 B_t(q_E^2)}{ A_t^2(q_E^2)q_E^2 + B_t^2(q_E^2) },\\
        A_b(p_E^2) =& 1 + \int^{\Lambda^2} d q_E^2 \mathcal{K}_{A_b}(p_E^2,q_E^2)\frac{A_b(q_E^2)}{ A_b^2(q_E^2)q_E^2 + B_b^2(q_E^2)}\\
        &+ \frac{{y_t}^2 + {y_b}^2}{32\pi^2}\int^{\Lambda^2} d q_E^2 K_A(p_E^2,q_E^2;\lambda_\phi \sigma_h^2)\frac{A_t(q_E^2)}{ A_t^2(q_E^2)q_E^2 + B_t^2(q_E^2) }\,, 
    \end{aligned}
\label{SD-b}
\end{equation}
where the integral kernels are given by
\begin{equation}
    \begin{aligned}
        &\mathcal{K}_{B_t}(x,y) = (1+\xi/3)\frac{3 {g_s}^2 C_F}{16\pi^2} K_B(x,y;0) + \frac{{y_t}^2}{32\pi^2} \left[ K_B(x,y;\lambda_\phi \sigma_h^2) - K_B(x,y;3\lambda_\phi \sigma_h^2) \right],\\
        &\mathcal{K}_{A_t}(x,y) = \frac{y}{x} \left\{ \xi\frac{4 {g_s}^2 C_F}{16\pi^2} K_A(x,y;0) + \frac{{y_t}^2}{32\pi^2} \left[ K_A(x,y;3\lambda_\phi \sigma_h^2)+K_A(x,y;\lambda_\phi \sigma_h^2) \right] \right\}\\
    \end{aligned}
\end{equation}
and
\begin{equation}
    \begin{aligned}
        K_B(x,y;M^2) &= \frac{2}{x + y + M^2 + \sqrt{(x + y + M^2)^2 - 4xy}},\\
        K_A(x,y;M^2) &= \frac{2xy}{\left[x + y + M^2 + \sqrt{(x + y + M^2)^2 - 4xy}\right]^2},
    \end{aligned}
\label{kernel}
\end{equation}
with $C_F = T^a T_a =  ({N_c}^2-1)/(2N_c) = 4/3$. We have regularized 
the momentum integrals by the momentum cutoff $\Lambda$. 
This cutoff scale $\Lambda$ acts as the ultraviolet regulator in the LSD equations, and is identified as infrared scales for the perturbative renormalization group running.
This is in the later section 
to coincide with a matching scale between the quark-meson model-like description at low energy and the present gauge-Higgs-Yukawa system.

\begin{figure}[H]
    \centering
    \includegraphics[width = 0.5\textwidth]{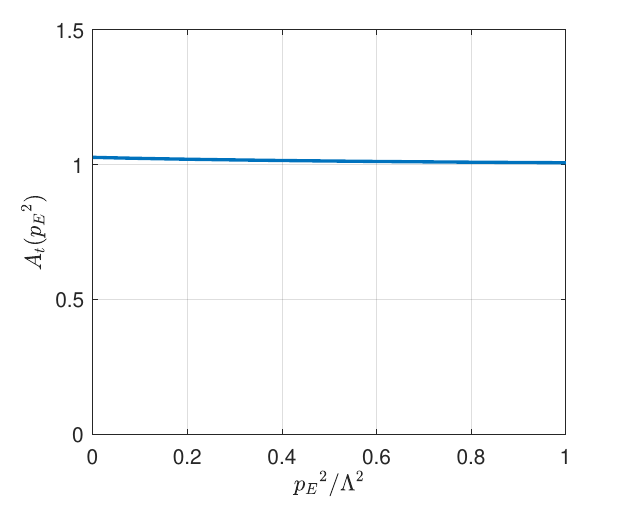}
    \caption{ The Euclidean momentum dependence of $A \equiv A_{t/b}$. The parameter setup has been taken as $y_t \simeq 2.6$, $g_s \simeq 3.1$, $y_b \simeq 0.063$, and $\lambda_\phi \simeq 0.93$ at the cutoff scale $\Lambda \simeq 234$ MeV below which the nonperturbative regime in the chiral broken phase governs.}  
    \label{fig:AvsP}
\end{figure}

Figure~\ref{fig:AvsP} shows the momentum dependence of $A_{t/b}(p_E^2)$ with $\xi=0$ in the regime that is what is called the nonperturbative regime, relevant to the present study with the input parameters to be addressed in details in the later section. 
We find $A_{t/b}(p_E^2) \simeq 1$ in the target momentum range,  thus 
the present extended ladder truncation works with $A_{t/b}=1$ in the Landau gauge $\xi=0$, 
in a way similar to the literature~\cite{Kondo:1993ty} in the $U(2)_L \times U(2)_R$ 
invariant limit.

\section{The CJT potential in the ladder approximation and phase structure of the gauge-Higgs-Yukawa system}

In this section,  
%, 
we work on the CJT formalism~\cite{Cornwall:1974vz} to get the CJT potential at the two-loop level in 
the ladder approximation and reproduce the LSD equations, Eqs.(\ref{SD-t}), (\ref{SD-b}), and (\ref{SD-HiggsVEV}). 
This CJT formalism also makes a preliminary setup to facilitate the matching procedure linked to a quark-meson model-like description in the later section. 
We then discuss the phase structure of the present gauge-Higgs-Yukawa system  
on the model parameter space along with the SM benchmark point, as referred to 
in Figs.~\ref{RG-SM} and~\ref{fig:AvsP}. 
The SM quark masses at the benchmark infrared scale ($\Lambda \simeq 234$ MeV) are also computed. 

\begin{figure}[t]
    \centering
    \includegraphics[width=0.35\textwidth]{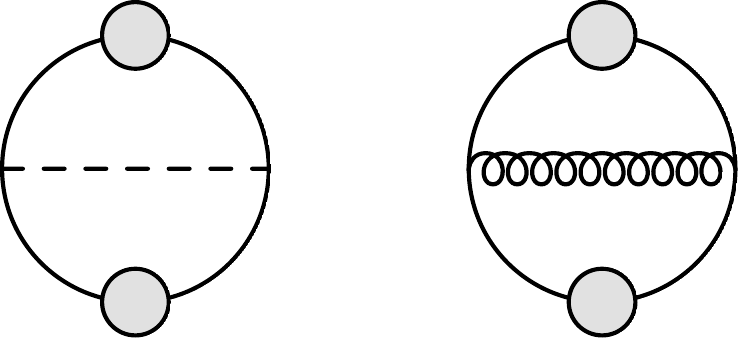}
    \caption{The two types of diagrams contributing to $V_2$ in Eq.(\ref{VCJT}) in the ladder approximation, where the solid curve represents quark propagators, the dashed one the scalars, and the wavy curve the one for the gluons. } 
\label{fig:Gamma2}
\end{figure}

\subsection{The CJT effective potential}

%It is also useful to derive the CJT potential of the top-Yukawa system in the bare vertex %approximation because we are going to match the scale anomaly in the latter section. 

We begin by writing the CJT effective potential 
in the presently applied ladder approximation up to the two-loop level, 
\begin{equation}
    V_{\rm eff}[\sigma_h, S_t, S_b] = V_0(\sigma_h)  
    + V_\text{scalar}^\text{1-loop}(\sigma_h, \Delta_{\sigma, \pi_i})
    +  
    V_\text{quarks}^\text{1-loop}(S_{t,b})
    + 
    V_\text{gauge}^\text{1-loop}(\Delta_g) 
     + V_2(\phi_c, \Delta_{\sigma, \pi_i}, S_{t, b}, \Delta_g) + \text{const.},
\label{VCJT}
\end{equation}
where $V_0(\sigma_h) = \frac{\lambda_\phi}{4}\sigma_h^4$ is the tree-level potential, and 
\begin{align}
  V_\text{scalar}^\text{1-loop}(\sigma_h,\Delta_{\sigma/\pi_i}) 
  &= - \frac{i}{2}\text{tr} \, \ln{ i\Delta_{\sigma_h/\pi_i}^{-1} } 
  %- \frac{i}{2}\text{tr} \left [ \mathcal{D}^{-1}_{\sigma, \pi_i}(\sigma_h)\Delta_{\sigma, \pi_i} \right ] 
  \,, \notag \\ 
  V_\text{quarks}^\text{1-loop}(S_{t/b}) 
  &=  i \, \text{tr} \, \ln{iS^{-1}_{t/b}} + i \, \text{tr} \left [ S^{-1}_{0 t/b}(\sigma_h)S_{t/b} \right ] \,, 
  \notag\\ 
  V_\text{gauge}^\text{1-loop}(\Delta_g) 
  & =  - \frac{i}{2} \text{tr} \, \ln{i\Delta_g^{-1}} 
  %- \frac{i}{2} \text{tr} \left [ \mathcal{D}^{-1}_g\cdot \Delta_g \right ] 
  \,, \label{V:1-loop}
\end{align}
with the inverse-tree propagators for $t$ and $b$ quarks defined as $S_{0 t/b}^{-1} 
= \slashed{p} - \frac{y_{t/b}}{\sqrt{2}} \sigma_h $. 
The $V_2$ term denotes the contribution from diagrams at the two-loop level, which 
in the present ladder approximation includes the graphs depicted in Fig.~\ref{fig:Gamma2}. 
In the first line of Eq.(\ref{V:1-loop}), 
$\Delta_{\sigma/\pi_i}^{-1}$ do not take into account 
the nonperturbative effects on the scalar two-point functions, i.e., $V_\text{scalar}^\text{1-loop}(\sigma_h,\Delta_{\sigma/\pi_i}) $ should be the one-particle irreducible potential for $\sigma$ and $\pi_i$ fields, which depends only on the Higgs VEV. 
Taking into account $A_{t/b}=1$ in the Landau gauge ($\xi=0$), one can compute $V_{\rm eff}$ to reach the form 
\begin{equation}
    \begin{aligned}
        V_\text{eff}[\sigma_h, B_t, B_b] 
        %= 
        %& \frac{\lambda_\phi}{4} {\sigma_h}^4 + \frac{1}{2} \int^\Lambda \frac{d^4 k_E}{(2\pi)^4} 
        %\ln{ \frac{k_E^2 + 3 \lambda_\phi \sigma_h^2}{k_E^2} } + \frac{3}{2} \int^\Lambda 
        %\frac{d^4 k_E}{(2\pi)^4} \ln{\frac{k_E^2 + \lambda_\phi \sigma_h^2}{k_E^2} } \\
        %&+ 4 N_c \int^\Lambda \frac{d^4 p_E}{(2\pi)^4}  \frac{B_t^2(p_E^2) -y_t \sigma_h {B_t}
        %(p_E^2)/\sqrt{2} }{p_E^2 + B_t^2(-p^2)} \\
        = &\frac{\lambda_\phi}{4} {\sigma_h}^4 + \frac{1}{32\pi^2} \int^{\Lambda^2} dk_E^2 \, k_E^2 \ln{ \frac{k_E^2 + 3 \lambda_\phi \sigma_h^2}{k_E^2} } + \frac{3}{32\pi^2} \int^{\Lambda^2} dk_E^2 \, k_E^2 \ln{\frac{k_E^2 + \lambda_\phi \sigma_h^2}{k_E^2} } \\
        &+ \frac{ N_c}{4\pi^2} \int^{\Lambda^2} d p_E^2 \, p_E^2  \frac{B_t^2(p_E^2) -y_t \sigma_h {B_t}(p_E^2)/\sqrt{2} }{p_E^2 + B_t^2(p_E^2)} 
        \notag\\ 
        & + \frac{ N_c}{4\pi^2} \int^{\Lambda^2} d p_E^2 \, p_E^2  \frac{B_b^2(p_E^2) -y_b \sigma_h {B_b}(p_E^2)/\sqrt{2} }{p_E^2 + B_b^2(p_E^2)}
        \,.
    \end{aligned}
    \label{Veff:CJT}
\end{equation}
The stationary conditions with respect to $B_t$ and $B_b$ respectively reproduce  
the SD equations for $B_t$ and $B_b$ in Eqs.(\ref{SD-t}) and Eq.(\ref{SD-b}) with the present ladder approximation prescribed.

The tadpole condition for the Higgs VEV in Eq.(\ref{SD-HiggsVEV}) is also derived 
with respect to $\sigma_h$, which, neglecting the small $b$ quark contribution, 
takes the same form as in Eq.(\ref{SD-HiggsVEV}). 
The three-point function in the left-hand side of Eq.(\ref{SD-HiggsVEV}) can be decomposed as 
\begin{equation}
    \braket{\sigma (\sigma^2  + \vec{\pi}^2)} = \braket{\sigma }^3 + 3\braket{\sigma }\braket{\sigma \sigma }_\text{conn.} + \braket{\sigma }\braket{\pi_i\pi_i}_\text{conn.} + \braket{\sigma (\sigma^2 + \pi_i^2)}_\text{conn.} 
    \,, \label{decomp}
\end{equation}
where the subscript ``conn." stands for the amplitude constructed from connected diagrams.   
The present ladder approximation the right-hand side of Eq.(\ref{decomp}) into only the first two terms with the bare bosonic propagators, so that we have 
\begin{equation}
    \begin{aligned}
       \braket{\sigma (\sigma^2  + \vec{\pi}^2)}  &= \lambda_\phi\sigma_h^3 + \lambda_\phi\sigma_h \int \frac{d^4q}{(2\pi)^4} \left( \frac{3i}{q^2-3\lambda_\phi\sigma_h^2} + \frac{3i}{q^2-\lambda_\phi\sigma_h^2} \right)\\ 
        &= \lambda_\phi\sigma_h^3 + \frac{3\lambda_\phi\sigma_h}{16\pi^2} \left( 2\Lambda^2 - 3\lambda_\phi\sigma_h^2\ln{\frac{\Lambda^2 - 3\lambda_\phi\sigma_h^2}{3\lambda_\phi\sigma_h^2}} - \lambda_\phi\sigma_h^2\ln{\frac{\Lambda^2 - \lambda_\phi\sigma_h^2}{\lambda_\phi\sigma_h^2}} \right).
    \end{aligned}
\end{equation}
The right hand side of Eq.(\ref{SD-HiggsVEV}), i.e., the top quark condensate $(- \frac{y_t}{\sqrt{2}} \langle \bar{t}t \rangle)$, 
is similarly simplified, in the present ladder approximation, to 
 \begin{equation}
    \begin{aligned}
     - \frac{y_t}{\sqrt{2}} \langle \bar{t}t \rangle &= \frac{y_t N_c}{\sqrt{2}}\int \frac{d^4q}{(2\pi)^4} \text{tr} S_t(q) = \frac{y_t N_c}{4\sqrt{2}\pi^2}\int^{\Lambda^2} dp_E^2 \frac{p_E^2B_t(p_E^2)}{p_E^2+{B_t^2(p_E^2)}}
     \,. 
    \end{aligned}
\end{equation}

The current quark masses are generated by the Higgs VEV $\sigma_h$ as 
the solution of Eq.(\ref{SD-HiggsVEV}), coupled with the SD equations for $B_t$ 
and $B_b$ in Eqs.(\ref{SD-t}) and (\ref{SD-b}), to be  
\begin{equation}
    m_q = \frac{y_q}{\sqrt{2}}\sigma_h,
\label{mq:current:def}
\end{equation}
for $q$-quark. 
The full masses for $t$ and $b$ quarks 
are defined as 
\begin{equation}
    m_{t/b,full} = B_{t/b}(p^2_E = m_{t/b,full}^2) 
    \,. \label{mtb:dyn}
\end{equation}

\subsection{The phase structure and implication to SM quark masses}

\begin{figure}[t]
    \centering
    \includegraphics[width=0.7\textwidth]{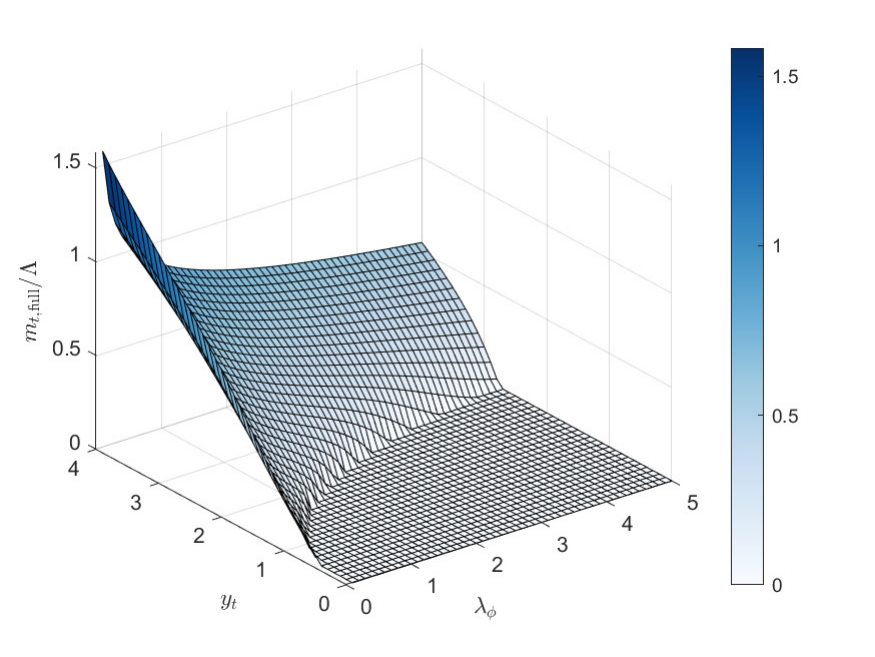}
    \caption{
    A phase diagram in the space spanned by 
the dynamical top mass normalized to the cutoff $m_{t, full}/\Lambda$, 
$y_t$, and $\lambda_\phi$, with $g_s = y_t$ and $y_b=0.5 y_t$. 
Here, though the isospin breaking by $y_b = 0.5 y_t$ has been set to be so small to be unrealistic, 
the generic feature of the phase diagram does not substantially alter for smaller $y_b$. 
}
\label{phase-diagram}
\end{figure}

Numerically solving the coupled LSD equations, Eqs.(\ref{SD-t}), (\ref{SD-t}), 
and (\ref{SD-HiggsVEV}), 
in Fig.~\ref{phase-diagram} we show a phase diagram in the space spanned by 
the dynamical top mass normalized to the cutoff $m_{t, full}/\Lambda$, 
$y_t$, and $\lambda_\phi$, with $g_s = y_t$ and $y_b=0.5 y_t$. 
The critical line has been observed on the $(y_t,\lambda_\phi)$ plane placed by 
$m_{t, full}=0$. In the limit where $\lambda_\phi \to 0$, $m_{t, full}$ goes divergent, 
which reflects the non-renormalizability of the Yukawa theory in the absence of the scalar quartic coupling. 

Here we have several remarks.  

\begin{figure}[t]
\centering
\label{(c)}
\includegraphics[width=0.6\textwidth]{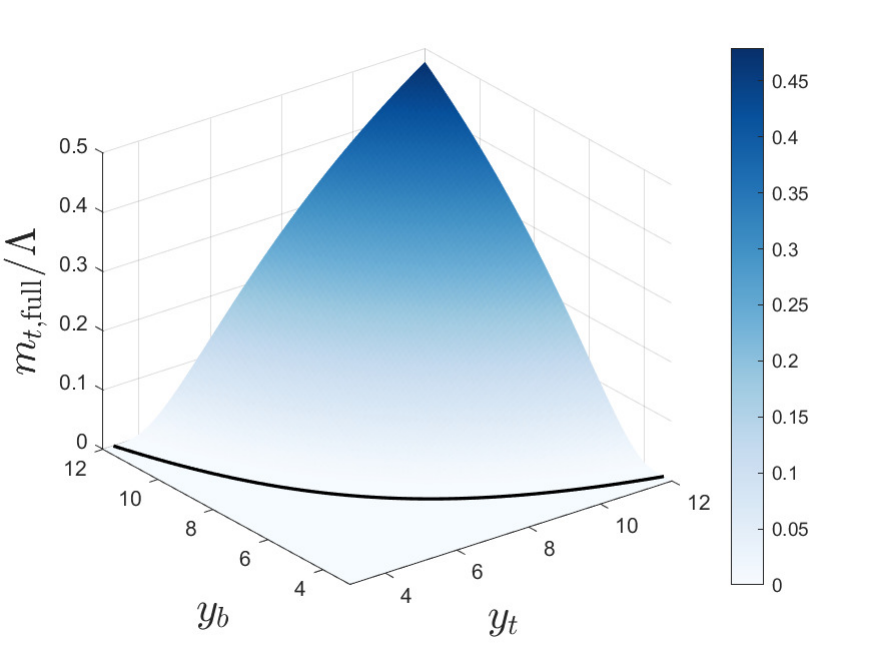}
\caption{ 
The phase diagram on 
the dynamical top mass normalized to the cutoff $m_{t, full}/\Lambda$}, 
$y_t$, and $y_b$, with $g_s = \lambda_\phi =0$. 
The critical line separating the chiral symmetric and broken phases is given as $y_t y_b = 4\pi^2$.  
\label{fig:Yukawa-limit}
\end{figure}

\begin{figure}[t]
\centering
\label{(c)}
\includegraphics[width=0.6\textwidth]{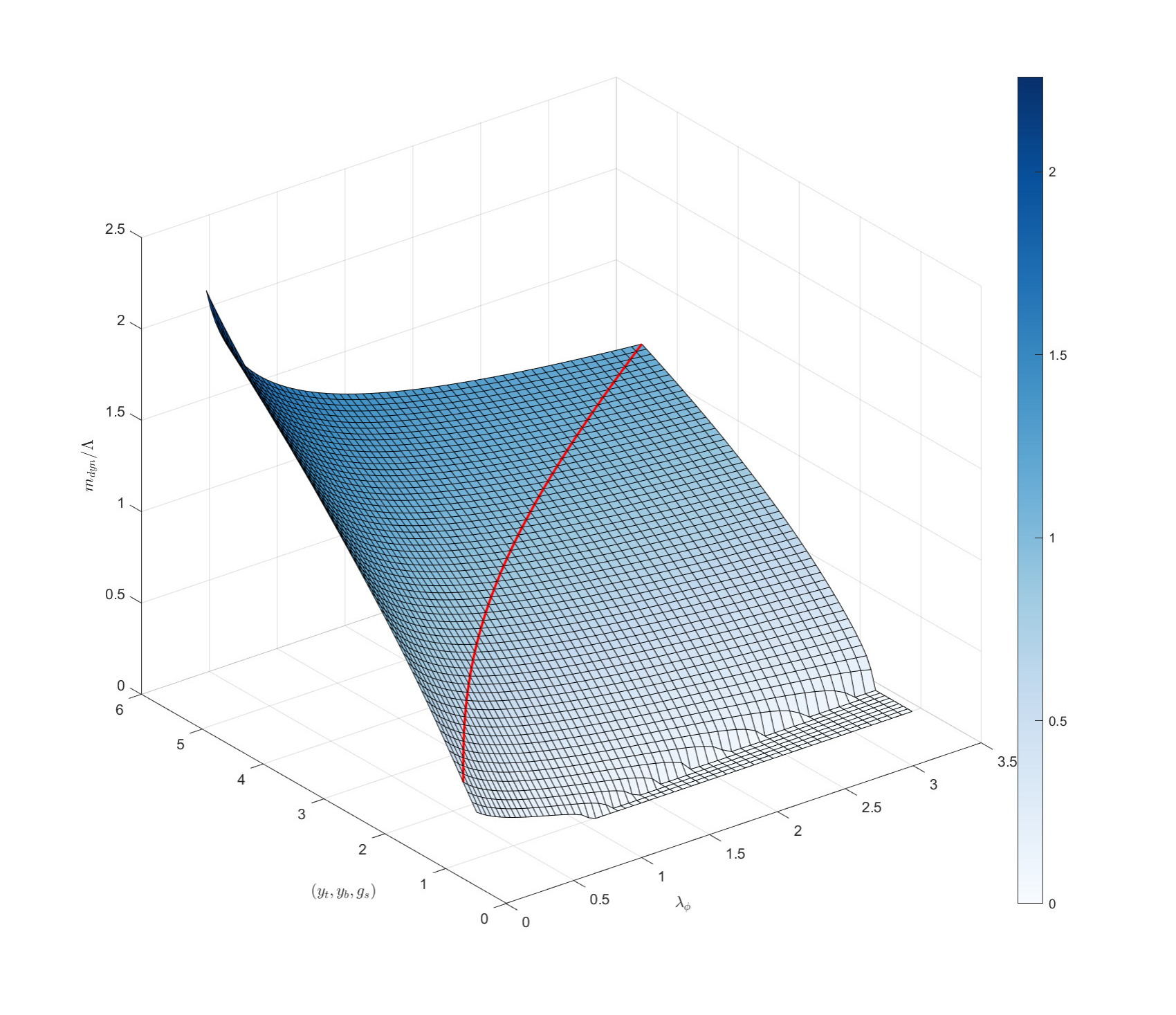}
\caption{ 
The phase diagram on 
the dynamical top mass normalized to the cutoff $m_{t, full}/\Lambda$, 
$(y_t, y_b, g_s)$ with respect to $\lambda_\phi$. 
The red curve seen in the center of the figure denotes 
the cutoff dependence trajectory of $m_{t, full}/\Lambda$ at the physical point with 
the realistic parameter setup, which is the same as 
the blue curve corresponding to the top full mass in Fig.~\ref{fig:full-masses}.  
}
\label{Phase_diagram-eps-converted-to}
\end{figure}

\begin{itemize}
    \item  
The bare parameter setup on $g_s=y_t$ well models the infrared feature of 
the renormalization group running in the SM as seen from Fig.~\ref{RG-SM}, when the cutoff $\Lambda$ for the SD equations matches with a strong coupling scale of $\alpha_s$, $\alpha_s(\Lambda) \sim \pi$.  

\item 
Though the input isospin breaking by $y_b = 0.5 y_t$ is too small to be realistic, the generic feature of the phase diagram does not substantially alter 
for smaller $y_b$. Just for a better illustration, we have chosen $y_b = 0.5 y_t$ 
in Fig.~\ref{phase-diagram} and also Fig.~\ref{fig:Yukawa-limit}. 
The realistic case such as $y_b = m_b/m_t y_t$ can be read off from another Fig.~\ref{Phase_diagram-eps-converted-to}.

\item 
The chiral symmetric phase is still observed even with nonzero $\lambda_\phi$ 
and $y_t$ as long as $y_t \lesssim 1$, 
because the tadpole of the Higgs is generated via the top quark condensate following the SD Eq.(\ref{SD-HiggsVEV}). 

\item 
In the strong QED limit where $y_t=y_b=0$ and $\lambda_\phi \to \infty$ (i.e. the decoupling limit of the Higgs), we have observed 
the well-known critical coupling $g_s^{\rm cr} =\pi $ ($\alpha_s = \pi/(3 C_F) = \pi/4)$)~\cite{Miransky:1994vk}. 

\end{itemize}

In the gaugeless limit with massless scalars, we have found a critical line $y_t y_b = 4\pi^2$ as shown in Fig.~\ref{fig:Yukawa-limit}. 
Note that in a conventional Yukawa theory with a single (pseudo)scalar coupled to a fermion bilinear, 
the critical line develops simply with the quadratic form of the Yukawa coupling, like 
$y_t^2 =$ constant.  
%where $y$ is associated with the effective charge among 
%the fermion scattering amplitude. 
However, the present case is more involved due to the presence of both 
$\sigma$ and $\pi$ exchanges: % in the corresponding scattering amplitude}:  
the dominant $y_t^2$ terms 
in the LSD equation for $B_t$ in 
Eq.(\ref{SD-t}) are actually canceled in the kernel function ${\cal K}_{B_t}$ in Eq.(\ref{kernel}) (with $\xi=0$).  
This is essentially due to an accidental $U(1)$ axial symmetry, between 
$\sigma$ and $\pi^3$ exchanges (repulsive and attractive channels),  
associated with 
the chiral rotation with respect to the third component of the Pauli matrix $\tau_3$. Under this chiral rotation, the complex Higgs field $\phi^0 (\sim  
\sigma + i \pi_3)$ and the quark doublet field $q$ in the Yukawa terms, in Eq.(\ref{GHY-Lag}) transform as
\begin{equation}
    \phi^0 \rightarrow e^{-i\theta} \phi^0 \quad \text{and} \quad q \rightarrow e^{i\frac{\theta}{2}\gamma_5\tau_3} q
\,.   \label{rota}
\end{equation}
When the EW charges get relevant, the $U(1)_A$ charges would be identical to the $U(1)_Y$ ones acting on 
$t$ and $\phi^0$. 
This symmetry protects the coupling strengths of $\bar{q}\sigma q$ and $\bar{q}\pi_3 q$ to be 
identical to each other, like $y_t (\bar{q} \sigma q + \bar{q} i \gamma_5 \pi_3 q)$. 
The cancellation between the $\sigma$ and $\pi_3$ exchange contributions is still approximately operative even  with the $U(1)_A$ violation due to 
the finite mass difference in $\sigma$ and $\pi_3$: 
${\cal K}_{B_t}(x,y;m^2) \simeq {\cal K}_{B_t}(x,y; 3 m^2)$ as long as 
$m/\Lambda \ll 1$~\footnote{
In the isospin symmetric limit with massless scalars  
we could have $y_t^2 = 4 \pi^2$. 
In this case, the cancellation will be due to the enhanced chiral $U(2)_L \times U(2)_R$ including the full four scalar and pseudoscalar exchanges. 
This case has been studied and discussed in the literature~\cite{Kondo:1993ty}, including 
this cancellation structure. 
}.

Thus, the strong top Yukawa coupling $y_t$ cannot solely develop 
the critical line or does not dominantly generate the chiral broken phase, or 
the top quark condensate. 
Therefore, the dynamical chiral symmetry in the present gauge-Higgs-Yukawa system is essentially triggered by the ladder QCD (i.e. strong QED), 
and the top flavor characteristics on the top-quark full mass $m_{t,full}$ 
is provided only through the overall top Yukawa coupling: 
$m_{t,full}$ still lies in the QCD scale $={\cal O}$(100) MeV.  
This conclusion favors the description of almost massless {\it six-flavor QCD} when the QCD phase transition is addressed, which is to be clarified right below. 
More discussions on the $U(1)_A$ symmetry leading to the conclusion above will be given 
later on, in Summary and Discussions.

\begin{figure}[t]
\centering
\subfigure[]{
\label{(a)}
\includegraphics[width=0.45\textwidth]{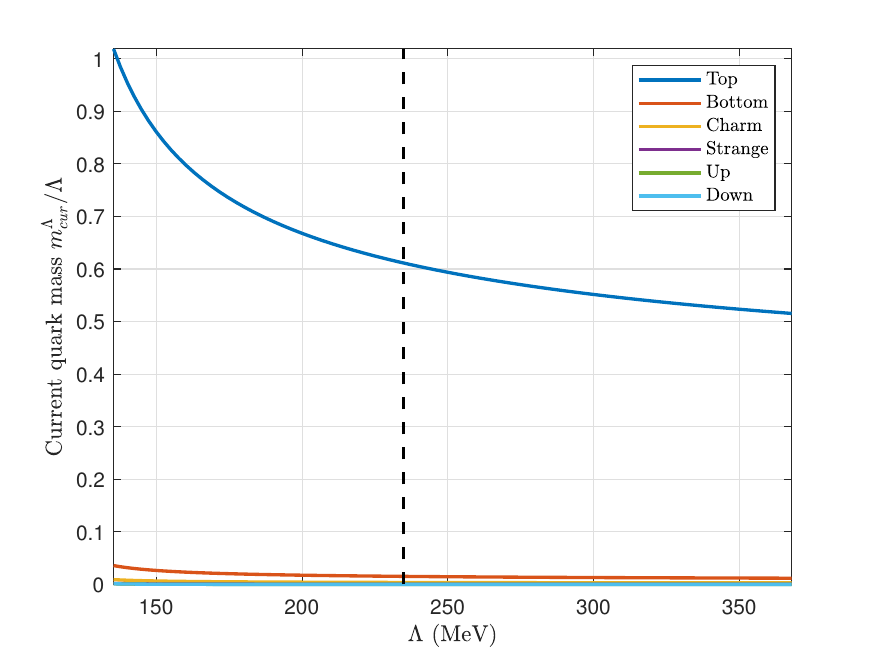}}
\quad
\subfigure[]{
\label{(b)}
\includegraphics[width=0.45\textwidth]{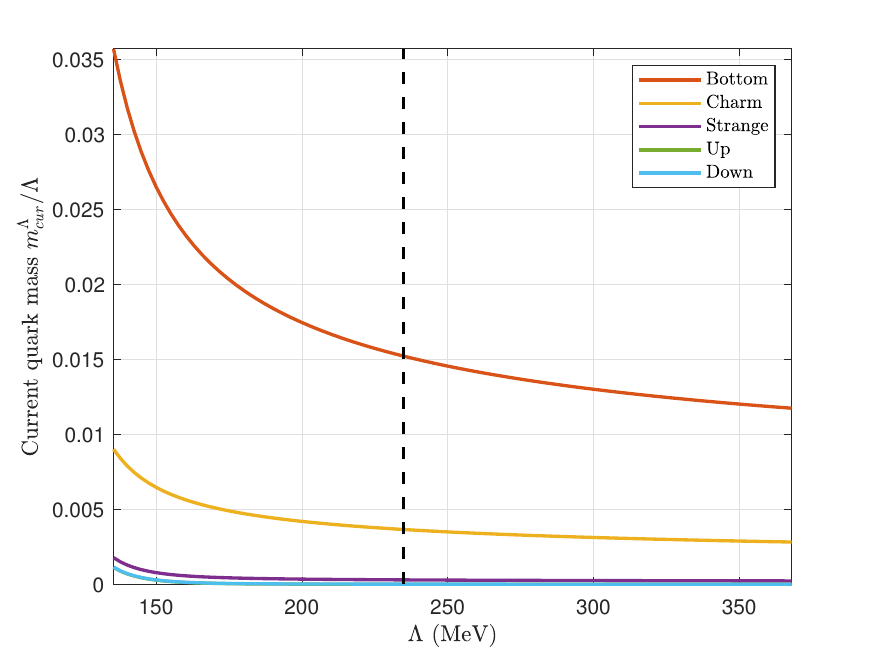}}
\caption{The plots of current quark masses versus $\Lambda$ including 
a reference point $\Lambda=234$ MeV,   
at which we have 
$g_s(\Lambda) \simeq 3.1 \sim \pi$, i.e., close to the critical coupling in strong QED, 
following the SM perturbative renormalization evolution as in Fig.~\ref{fig:AvsP}. 
The panel (b) excludes the top quark, while the panel (a) includes it. }  
\label{fig:current-masses}
\end{figure}

Figure~\ref{fig:current-masses} shows 
the current quark masses as a function of $\Lambda$, which arise from the Higgs tadpole contribution in Eq.(\ref{mq:current}) together with the LSD solutions. 
The selected range of $\Lambda$ includes  
a reference point $\Lambda=234$ MeV where 
we have $g_s(\Lambda) \simeq 3.1 \sim \pi$, i.e., close to the critical coupling in strong QED, 
which corresponds to a boundary between the perturbative 
nonperturbative coupling regimes 
as quoted in Fig.~\ref{fig:AvsP}: 
at $\Lambda=234$ MeV the SM perturbative two-loop running, 
with the EW scale 
inputs as in Fig.~\ref{RG-SM}, yields 
$y_t \simeq 2.6$, $g_s \simeq 3.1$, $y_b \simeq 0.063$, and $\lambda_\phi \simeq 0.93$. 
Figure~\ref{fig:AvsP} indeed implies that at around the reference point 
$\Lambda=234$ MeV, we have current quark masses for the SM heavier three quarks 
(the Higgs tadpole contribution in Eq.(\ref{mq:current})),  
\begin{align}
    m_t(\Lambda=234\,{\rm MeV}) &\simeq 140 \,{\rm MeV} 
    \,, \notag\\ 
    m_b(\Lambda=234\,{\rm MeV}) &\simeq 3.5 \,{\rm MeV} 
    \,, \notag\\ 
    m_c(\Lambda=234\,{\rm MeV}) &\simeq 0.94 \,{\rm MeV} 
    \,. \label{mq:current}
\end{align}
In addition, the top and bottom quarks get the pure ladder QCD and Yukawa contributions due to the top condensation. At the same reference 
point of $\Lambda$, we find the full masses for $t$ and $b$ quarks,  
\begin{align}
    m_{t, full}(\Lambda=234\,{\rm MeV}) &\simeq 175 \, {\rm MeV} 
    \,, \notag\\ 
    m_{b, full}(\Lambda=234\,{\rm MeV}) &\simeq  12 \, {\rm MeV} 
\,. \label{mdyn:tb}
\end{align}
Thus all the six quarks can be light enough in the typical QCD phase transition epoch at around the temperature $\sim 100 - 200$ MeV.

\begin{figure}[t]
\centering
\subfigure[]{
\label{(a)}
\includegraphics[width=0.45\textwidth]{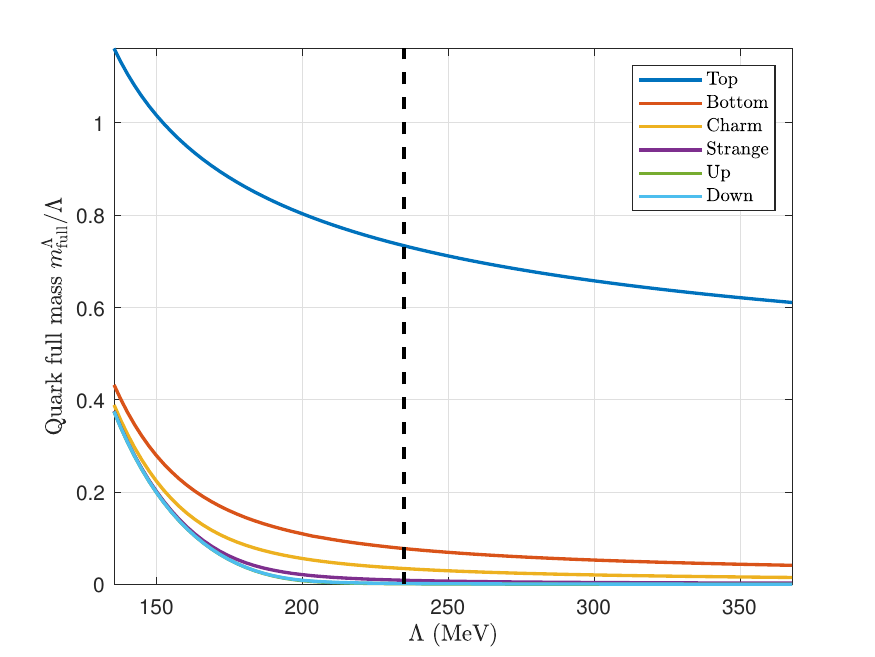}}
\quad
\subfigure[]{
\label{(b)}
\includegraphics[width=0.45\textwidth]{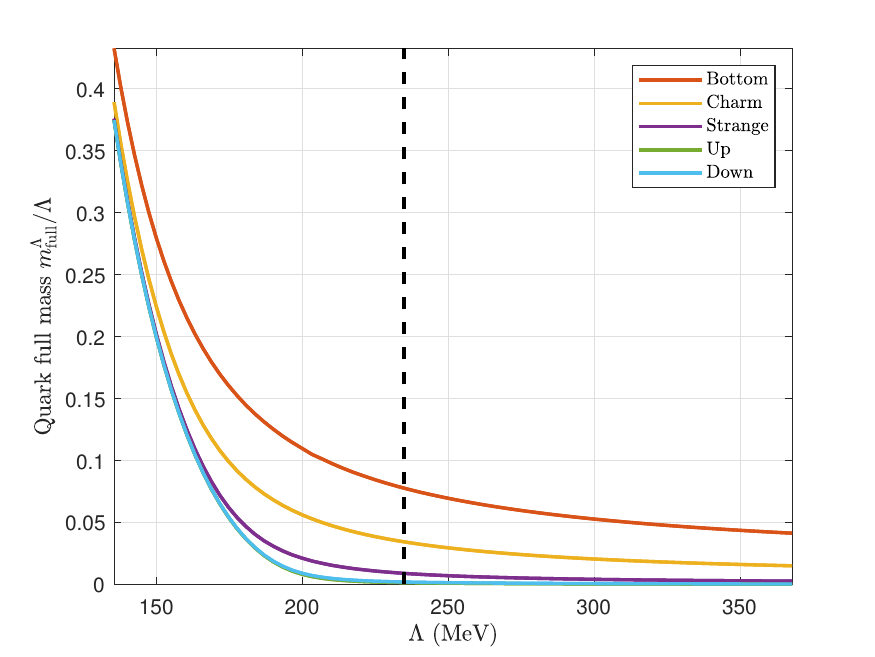}}
\caption{Quark full masses as a function of $\Lambda$ in the same range as in Fig.~\ref{fig:current-masses}. }
\label{fig:full-masses}
\end{figure}

\section{Cosmological implication of the top quark condensation in supercooled EW phase transition}

Since the dynamically generated top condensate couples to the SM Higgs, the supercooled EW phase transition should get a significant correlation 
with the thermal evolution of the top condensate. 
In this section, we address this issue by employing a quark-meson model 
in the large $N_c$ limit. namely, the mean-field approximation for 
the Higgs and QCD meson fields. 
We fix the model parameters by matching them to the LSD results with 
the matching scale set at 
$\Lambda = 234$ MeV, which is identified as the LSD cutoff scale as has been 
done in the previous section. 
We then discuss the cosmological phase transition arising from the evidence 
of the top quark condensation and its cosmological consequences such as 
the GW production. 

\subsection{Quark meson model for the top-Higgs sector}

%\subsubsection{Effective degree of freedom}

\begin{figure}[t]
    \centering
    \includegraphics{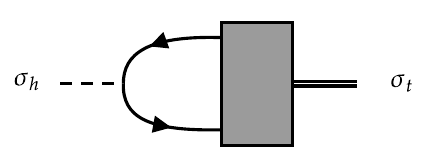}
    \caption{The Feynman graph generating the $\sigma_t$ - $\sigma_h$ mixing. 
    The blob and (internal) top loop lines include the nonperturbative ladder-type contributions as has been discussed in Sec.III. } 
    \label{fig:t-h-mix}
\end{figure}

We consider a quark-meson model-like model for the low-energy 
effective description for the present gauge-Higgs-Yukawa system 
at the matching scale $\Lambda$. 
Since the correlation with the SM Higgs arising only through the Yukawa couplings to quarks, the mean-field associated with only the top quark condensate becomes significant. 
The model Lagrangian is thus constructed from 
the scale-invariant SM Higgs sector (only with neutral scalars) coupled to 
the top quark:  
\begin{align}
        \mathcal{L}_\text{QMM} 
        &= \bar{t} i \slashed{\partial} t - g \, \bar{t}_{L} M_{\bar{t}t} t_{R} - y_t \, \bar{t}_{L} M_h t_{R}  + \text{h.c.} 
        \notag \\
        &+  \partial_\mu M_{\bar{t}t}^\dag  \partial^\mu M_{\bar{t}t} 
+  \partial_\mu M_h^\dag \partial^\mu M_h  - V_{\rm QMM}
\,, 
\notag\\  
V_{\rm QMM} 
&= \lambda_\phi \left( M_h^\dagger M_h \right)^2 
 - g_m (M_h^\dag M_{\bar{t}t} + {\rm h.c.}) 
        + V(M_{\bar{t}t})
        \,, 
\label{L:QMM}
\end{align}
where $M_{\bar{t}t} = (\sigma_t + i \pi_t)/\sqrt{2}$ for the $\bar{t}t$ scalar meson, i.e., topponium, and $M_{h} = (\sigma_h + i \pi_h)/\sqrt{2}$ for 
the neutral Higgs scalar fields in the SM. 
Here several comments are in order: 

\begin{itemize} 
\item 
The $g_m$ coupling term denotes the mixing between the topponium and the SM Higgs mean fields,
$- g_m \langle \sigma_h \rangle \langle \sigma_t \rangle$, 
which corresponds to the disconnected graph contribution in the ladder approximation. 
See Fig.~\ref{fig:t-h-mix} for the typical Feyman graph.   
The figure also displays a squared blob, which denotes the Bethe-Salpeter amplitude to describe the dynamical $\sigma_t$ - $t$ - $t$ vertex, including not merely the static 
potential terms, but also the momentum dependence. The latter type of quantum corrections are discarded in the current ladder approximation. 
The momentum dependence will not give any 
significance on the static potential analysis aiming at discussion on 
the thermal phase transition, which might, however, be the issue when 
the dynamic cosmological phase transition is addressed. We will come back 
to this point later in the Summary and Discussion section.

\item 
In addition to the tadpole term, in Eq.(\ref{L:QMM}) the top Yukawa coupling to $M_h$ has been taken into account, so that the  top-quark full mass contributions are shared by those two terms due to the presence of the dynamical Higgs field. 
This is in contrast to 
the case of the ordinary QCD quark-meson model which includes the current quark mass effect in the tadpole term and the dynamical mass part in the Yukawa term.

\item 
One can also see that the second derivative of the CJT potential with respect to 
$\sigma_h$ in Eq.(\ref{Veff:CJT}) vanishes, which implies no Higgs portal 
coupling to $\sigma_t$, such as $|M_h|^2 |M_{\bar{t} t}|^2$, 
hence the $g_m$ coupling is the unique way to allow 
the mixing of $M_{\bar{t}t}$ with $M_h$ up to operators smaller than 
dimension four.

\item 
The topponium VEV at the vacuum, $\braket{\sigma_{t}}$, is set by the topped pion decay constant 
$f_t$, which we nonperturbatively compute 
by using the Pagels-Stokar formula~\cite{Pagels:1979hd} with the solution of the coupled LSD equations (Eqs.(\ref{SD-t}), (\ref{SD-b}), and (\ref{SD-HiggsVEV})) 
as  
\begin{equation}
    f_t^2 = \frac{N_c}{4\pi^2} \int^{\Lambda^2}  dp_E^2 \cdot p_E^2  \frac{B_t^2(p_E^2) - \frac{p_E^2}{4}\frac{d}{dp_E^2}B_t^2(p_E^2) }{(p_E^2+B_t^2(p_E^2))^2}\,. 
\label{ft}
\end{equation}
The numerical estimate gives $f_t/\Lambda \simeq 0.091$ and for Higgs VEV $\sigma_h/\Lambda \simeq 0.34$ at $\Lambda=234$ MeV with the same inputs as in Eqs.(\ref{mq:current}) and (\ref{mdyn:tb}).

\end{itemize}

Besides, we have the potential term for $M_{\bar{t}t}$ in Eq.(\ref{L:QMM}). 
Taking $\pi_t =0$ via the U(1) axial transformation, 
the mesonic potential can be written as a function of only $\sigma_t$. 
At this point, a simple-minded parametrization of the potential is of the Ginzburg-Landau polynomial form, 
\begin{equation}
    V(M_{\bar{t}t}) \Bigg|_{\rm GL} = \mu^2 \sigma_t^2 + \lambda \sigma_t^2 
    \,. \label{GL-type}
\end{equation}
The presence of the mass term $\mu^2$ would reflect the hard scale breaking at the deeper infrared regime of QCD, 
simply following the perturbative running of the gauge coupling (as seen also from 
Fig.~\ref{RG-SM}). 
Still, we could have another possibility featuring the deeply infrared QCD nature: 
in a recent nonperturbative analysis based on the 
functional renormalization group, like in the literature~\cite{Deur:2023dzc}, 
the QCD gauge coupling might possess a nontrivial infrared fixed point, 
which implies an almost scale invariance (to be eventually broken by the dynamical chiral symmetry breaking, though)~\footnote{
The almost scale-invariant QCD with six quarks has also been discussed in works 
other than the functional renormalization group analysis, motivated by 
addressing the QCD sigma meson as a dilaton associated with the spontaneous breaking of the approximate scale invariance~\cite{Li:2016uzn,Cata:2019edh,Zwicky:2023bzk,Zwicky:2023krx,Crewther:2020tgd,Crewther:2013vea}. }. 
From this perspective, we might also consider the Coleman-Weinberg type~\cite{Coleman:1973jx},  
\begin{equation}
    V(M_{\bar{t}t}) \Bigg|_{\rm CW}  = \lambda \sigma_t^4 \left[ a + b \ln\left( \frac{\sigma_t}{f_t} \right) \right]
    \,. \label{CW-type}
\end{equation}
Taking into account both two possibilities, 
thus we model the form of $V(M_{\bar{t}t})$ as 
\begin{equation}
    V_{M_{\bar{t}t}} = \lambda_D \, 
    \left( \frac{\sigma_t}{f_t} \right)^{4-\gamma_D} {f_t}^4 +\lambda \left( \frac{\sigma_t}{f_t} \right)^{4} {f_t}^4.
\end{equation}
Here we have introduced an anomalous dimension parameter $\gamma_D$ associated with 
the scale anomaly, which can be fixed by solving the underlying theory. 
In the present study, we  
take it as a free parameter and consider $\gamma_D$ in a range 
of $0 < \gamma_D \le 2$. 
The $\lambda_D$ term thus plays a role of 
interplay between the Ginzburg-Landau and almost scale-invariant 
description.  
When $\gamma_D = 2$ the $\lambda_D$ term is reduced to the normal quadratic mass term as in Eq.(\ref{GL-type}), while for $\gamma_D \simeq 0$ it provides a logarithmic potential term like 
$\lambda_D \, \gamma_D  \sigma_t^4  \log (\sigma_t/f_t)$:  
\begin{align}
        V(M_{\bar{t}t}) \Bigg|_{\gamma_D \simeq 0} 
        %&= \lambda_D\left( \frac{\sigma_t}{f_t} \right)^{4} e^{-\gamma_D \ln\left( 
        %\frac{\sigma_t}{f_t} \right)} {f_t}^4 +\lambda \left( \frac{\sigma_t}{f_t} 
        %\right)^{4} {f_t}^4 \notag \\
        &= \sigma_t^4 \left[ (\lambda+\lambda_D) - \lambda_D \gamma_D \ln\left( \frac{\sigma_t}{f_t} \right) + \mathcal{O}(\gamma_D^2) \right] 
        \equiv V(M_{\bar{t}t}) \Bigg|_{\rm CW} \quad {\rm in} \quad {\rm Eq.(\ref{CW-type})}
     \,. 
\end{align}

We fix the vacuum feature of the model including the Higgs 
VEV $\langle \sigma_h \rangle$, the topponium VEV $\langle \sigma_t \rangle = f_t$, and 
the potential energy i.e., the scale anomaly, 
by matching with the 
nonperturbative results associated with the CJT effective potential 
in Eq.(\ref{Veff:CJT}). 
First, the stationary condition for the potential terms, together with 
the LSD solutions to $\langle \sigma_t \rangle = f_t$ (Eq.(\ref{ft})) and $\langle 
\sigma_h \rangle$ (Eq.(\ref{SD-HiggsVEV})), constrain 
$\lambda_D$, $\gamma_D$, and $g_m$, as follows: 
\begin{equation}
    \begin{aligned}
        & \left. \frac{\partial V}{\partial \sigma_h} \right|_\text{vacuum} = \lambda_\phi \sigma_h^3 - g_m f_t = 0;\\
        & \left. \frac{\partial V}{\partial \sigma_t} \right|_\text{vacuum} = \lambda_D (4-\gamma_D) {f_t}^3 + 4 \lambda {f_t}^3 -g_m \sigma_h = 0.
    \end{aligned}
\label{stcond}
\end{equation}
Then, matching with the vacuum energy gives a further constraint like 
\begin{equation}
        V_{\rm CJT}\Bigg|_{\rm vacuum} \equiv  
        V_{QMM}\Bigg|_{\rm vacuum} 
        = \frac{1}{4} \left(\lambda_D {f_t}^4 - 2 g_m \langle \sigma_h \rangle f_t  \right) 
        %= \frac{1}{4} \partial_\mu D^\mu, 
        \,. 
\end{equation}

Though the $g$ coupling between quark and meson is at this point just free parameter,   referring to the LSD results may fix it via the constituent quark mass in the following two ways (I and II) 
\begin{equation}
\begin{aligned}
    &g|_I = \left(m_{t, full} -  \frac{y_t}{\sqrt{2}} \langle  \sigma_h \rangle \right) / f_t,\\
    &g|_{II} = m_{t, full} / f_t.
\end{aligned}
\end{equation}
In the former case ($g|_I$), $M_{\bar{t}t}$ couples to the top quark pair only via 
the dynamical mass of the top quark, while keeping the tadpole term coupled to the Higgs mean field. The latter case ($g|_{II}$) implies that the top quark pair still couples to 
$M_{\bar{t}t}$ also through the current mass, so that there are two current mass contributions to the model along with the tadpole term.  
The LSD result at the reference point $\Lambda=234$ MeV 
gives 
\begin{align}
    g|_I & \simeq  1.3 
    \,, \notag\\ 
    g|_{II} & \simeq  8.0
\,. 
\label{g1-2}
\end{align}

Thus the model parameters can completely be fixed once  $\gamma_D$ is given for 
$0 < \gamma_D \le 2$. 
Table~\ref{tab:para} gives the model parameter sets fixed by matching with 
the present LSD analysis as described above. 

\begin{table}[H]
\centering
\begin{tabular}{c|cccccc}
\hline
$\gamma_D$      & \multicolumn{2}{c|}{2}                                    & \multicolumn{2}{c|}{0.5}                                  & \multicolumn{2}{c}{0.1}              \\ \hline
$g$             & \multicolumn{1}{c|}{1.3} & \multicolumn{1}{c|}{8.0} & \multicolumn{1}{c|}{1.3} & \multicolumn{1}{c|}{8.0} & \multicolumn{1}{c|}{1.3} & 8.0 \\ \hline
$g_m/\Lambda^2$ & \multicolumn{6}{c}{0.39}                                                                                                                                   \\ \hline
$\lambda$       & \multicolumn{2}{c|}{50}                              & \multicolumn{2}{c|}{89}                              & \multicolumn{2}{c}{299}         \\ \hline
$\lambda_D$     & \multicolumn{2}{c|}{-13}                             & \multicolumn{2}{c|}{-52}                             & \multicolumn{2}{c}{-262}        \\ \hline
\end{tabular}
\caption{The parameter setting of the present quark-meson model fixed by the matching with the LSD analysis at $\Lambda = 234$ MeV,  
given a value of $\gamma_D$ 
in a range of $0< \gamma_D \le 2$. } 
\label{tab:para}
\end{table}

\subsection{Thermal phase transition} 

We incorporate the thermal loop corrections arising from 
the top-Yukawa interaction at the one-loop level of the quark-meson model in Eq.(\ref{L:QMM}), based on 
the imaginary time formalism, to derive the thermodynamic potential ($\Omega(\sigma_t, \sigma_h)$)
for the mean fields $\sigma_t$ and $\sigma_h$. 
The top loop corrections to the vacuum part are not included in 
the thermodynamic potential, instead, which involves nonperturbative 
corrections from the LSD analysis in the tree level as has been discussed above.

\begin{figure}[t]
\centering
\subfigure[]{
\label{(a)}
\includegraphics[width=0.4\textwidth]{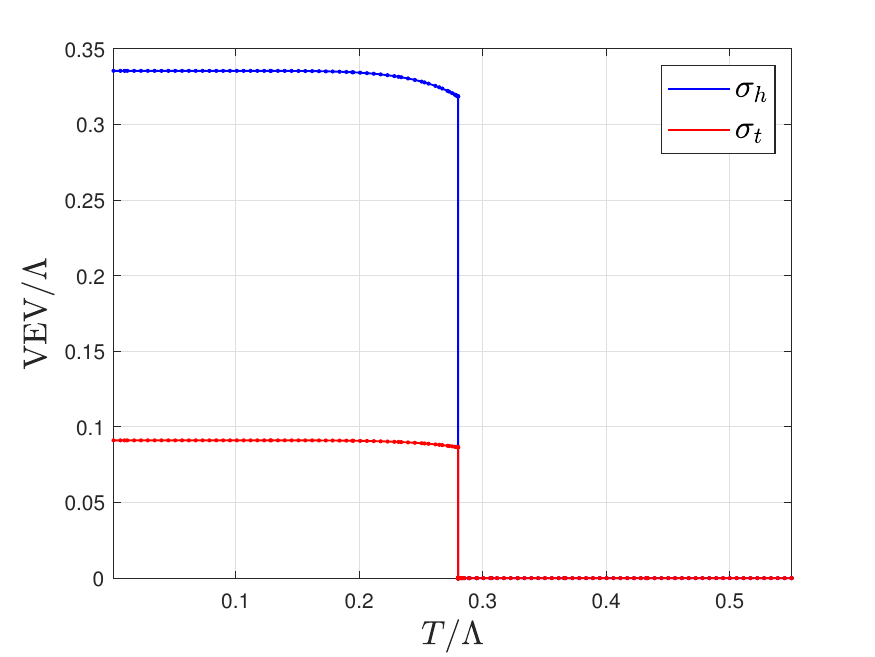}}
\quad 
\subfigure[]{
\label{(b)}
\includegraphics[width=0.4\textwidth]{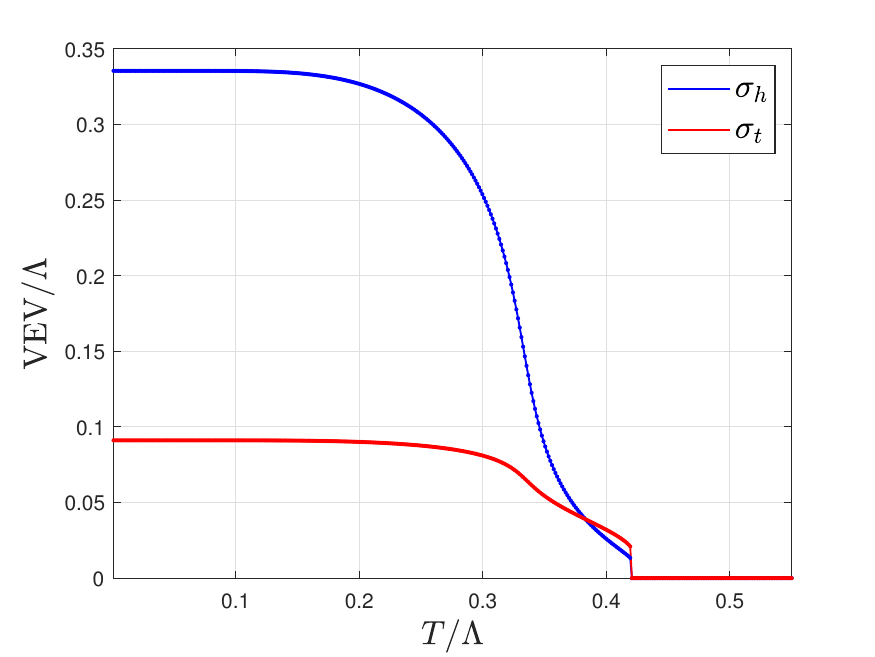}}
\caption{The temperature dependence of $\sigma_t$ and $\sigma_h$ in the case of $\gamma_D = 2$ for (a) $g|_{II} \simeq 8.0$ and (b) $g|_{I} \simeq 1.3$.}
\label{fig:T-sigma}
\end{figure}

\begin{figure}[H]
\centering
\subfigure[]{
\label{(a)}
\includegraphics[width=0.3\textwidth]{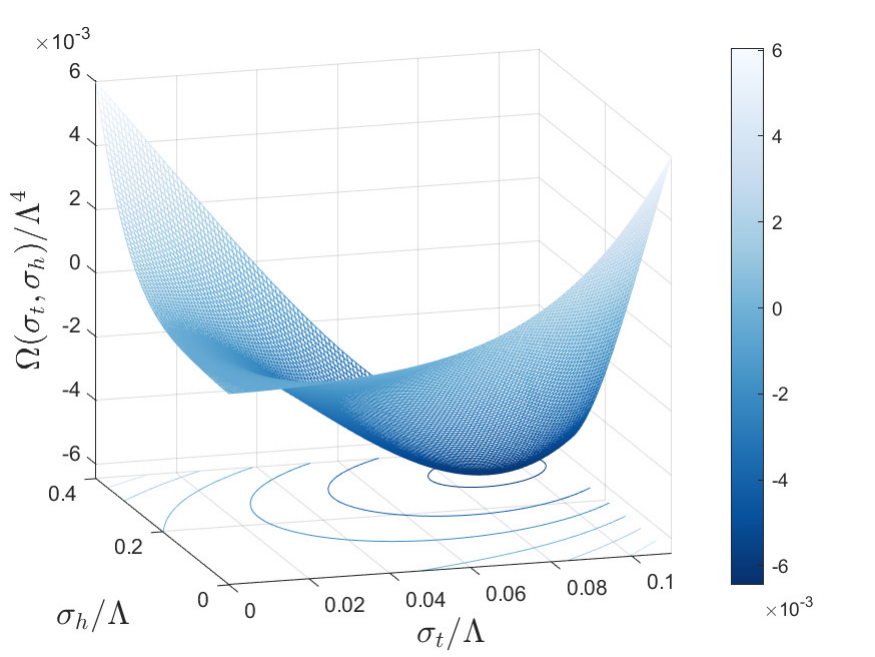}}
\quad
\subfigure[]{
\label{(b)}
\includegraphics[width=0.3\textwidth]{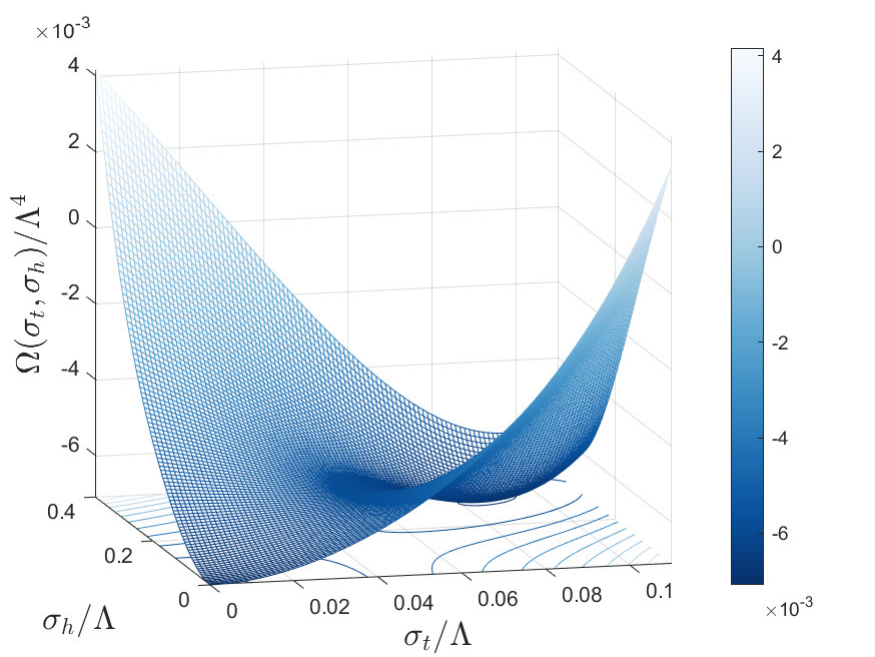}}
\subfigure[]{
\label{(c)}
\includegraphics[width=0.3\textwidth]{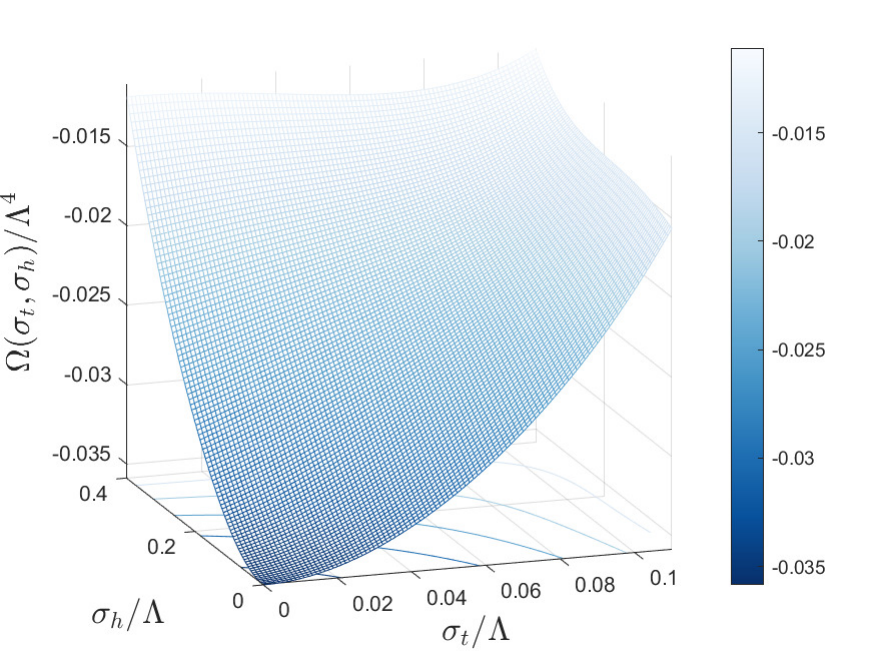}}
\caption{The thermodynamic potentials as a function of $\sigma_t$ and $\sigma_h$ for
$\gamma_D = 2$ with $g|_{II} \simeq 8.0$ around the first order phase transition point. 
The panel (a) corresponds to the case when $T = 0.5 T_c$; 
(b) $T = 1.0 T_c$; (c) $T = 1.5 T_c$, where $T_c \simeq 65$ MeV 
for the matching scale $\Lambda=234$ MeV.  }
\label{fig:Omega}
\end{figure}

Figure~\ref{fig:T-sigma} shows the temperature ($T$) dependence (normalized to 
$\Lambda=234$ MeV) 
of 
$\sigma_t$ and $\sigma_h$ in the case of $\gamma_D = 2$ with 
$g|_I$ and $g|_{II}$ in Eq.(\ref{g1-2}). 
For larger $g$, the phase transition goes like the type of the first-order 
with the critical temperature $T_c \simeq 65$ MeV, while 
for smaller $g$, it tends to be changed to crossover with the pseudocritical 
temperature (defined as the inflection point of the order parameter) $T_{\rm pc}\simeq 80$ MeV, which holds approximately for both $\sigma_t$ and $\sigma_h$. 
We have checked that this trend does not substantially change 
for smaller $\gamma_D$. 
The first-order phase transition has been undergone due to the emergence of 
the thermal potential barrier between the false and true vacua, which 
has been created along 
a hybrid direction involving two nonzero $\sigma_t$ and $\sigma_h$, 
as plotted in Fig.~\ref{fig:Omega}. 
This is the characteristic consequence of the presence of the top quark condensate 
in the thermal history.

\subsection{Cosmological phase transition}

%For the first-order phase transition, there is an out-of-equilibrium effect called 
%supercooling, which describes the false vacuum decay. So far as I know, there are
%two approaches: the conventional Euclidean approach\cite{Callan:1977pt} and the
%real-time false vacuum decay approach\cite{Batini:2023zpi}. In this work we follow
% the usual way, applying the classical bounce solution. Through the general
%background, we assume that the Universe gets a Hubble background similar to the
%reference\cite{Sagunski:2023ynd}.

The first-order phase transition in the hybrid $\sigma_t$-$\sigma_h$ system 
at around $T_c \simeq 65$ MeV would be supercooled in the universe to produce the bubble nucleation and percolation. 
We discuss this supercooling in terms of the false vacuum decay and 
evaluate the false vacuum decay rate approximately as~\cite{Linde:1981zj}
\begin{equation}
    \Gamma \simeq T^4 \left( \frac{\mathcal{S}_3}{2\pi T} \right)^\frac{3}{2} \exp{\left(- \frac{\mathcal{S}_3}{T} \right)}\,, 
\end{equation}
where the $\mathcal{S}_3 = \mathcal{S}_3[\phi_\text{b}(r)]$ is the three-dimensional ${\cal O}(3)$ symmetric Euclidean action~\footnote{
As discussed in Ref.~\cite{Helmboldt:2019pan}, the quark-meson model introduces the mesonic degrees of freedom as {\it fundamental},  $\mathcal{S}_3$ takes the form having the canonical kinetic term for the meson fields. 
},  
\begin{equation}
    \mathcal{S}_3 = \int dr \, (4\pi r^2) \left[ \frac{1}{2} \left( \frac{d\vec{\sigma}}{dr} + \Omega(\vec{\sigma},T) \right) \right]
\,,\label{BA}
\end{equation}
with a set of the bounce solutions $\vec{\sigma}(r) = (\sigma_t,\sigma_h)^T$. 
The bounce solutions are given by solving the stationary condition of the action: 
\begin{equation}
    \frac{d^2 \vec{\sigma}}{dr^2} + \frac{2}{r} \frac{d\vec{\sigma}}{dr} = \nabla_\sigma \Omega(\vec{\sigma}, r) \,, 
\end{equation}
with the boundary conditions
\begin{equation}
    \begin{aligned}
        &\left. \frac{d\vec{\sigma}}{dr} \right|_{r = 0} = 0,\\
        &\left. \vec{\sigma} \right|_{r \rightarrow \infty} = 0.
    \end{aligned}
\end{equation}
To get the multi-field bounce solutions, we have used the numerical source pack \texttt{CosmoTransitions}~\cite{Wainwright:2011kj}. 

%Since we are considering the second scenario of the EW supercooling, we can assume the Universe has a Hubble background $H^2(T_i) = \frac{\pi^2}{90}g_*\frac{{T_i}^4}{{M_\text{Pl}}^2}$ induced by the hidden sector and characteristic it by introducing the "total reheating temperature" $T_i$ since the QCD phase transition triggers the EWPT as in the reference\cite{Sagunski:2023ynd}. In principle, we can also consider the case that the EW supercooling does not exit after the QCD phase transition. In this case, the Hubble parameter is dominated by the radiation and the vacuum energy
Then the nucleation temperature $T_n$ can be determined by
\begin{equation}
    \frac{\Gamma(T_n)}{H^4(T_n)} \sim  1.
\end{equation}
where the Hubble parameter is parametrized as 
\begin{equation}
    H^2(T) = \frac{\rho_\text{rad}(T) + \rho_\text{vac}(T)}{{3M_\text{Pl}}^2} \simeq  \frac{1}{{3M_\text{Pl}}^2} \left( \frac{\pi^2}{30} g_*(T_i) T_i^4 + \Delta V \right)\,, 
\end{equation} 
with $\Delta V$ being the vacuum energy density difference between the false vacuum and the true vacuum. 
Here $g_*(T)$ denotes the effective degrees of freedom at $T$, and 
$M_{\rm pl}$ is the reduced Planck scale $\simeq 1.2 \times 10^{18}$ GeV. 
We have also taken into account the dark-sector induced potential density, which is parametrized by a dark thermal density with $T_i$ as done in the literature~\cite{Sagunski:2023ynd}. 
This term can be turned off or on depending on whether the electroweak/dark sector phase transition is triggered before the hybrid $\sigma_t$ - $\sigma_h$ phase 
transition takes place, or not.

We also evaluate the percolation temperature $T_p$ via 
%as has been studied in the %reference\cite{Ellis:2018mja}. 
%We consider 
the probability still staying in the false vacuum~\cite{Guth:1979bh,Guth:1981uk}, 
%\begin{equation}
$    P = e^{-I(T)}$,   
%\end{equation}
where the exponent function is available in the literature~\cite{Ellis:2018mja,Sagunski:2023ynd} 
%\begin{equation}
%    I(T) = \frac{4\pi}{3} \int_{T}^{T_c} \frac{d\prime{T} \, \Gamma(\prime{T})}
%%{H(\prime{T})\prime{T}^4} \left( \int_{T}^{\prime{T}}d\tilde{T} \frac{v_w}
%{H(\tilde{T})} \right)^3.
%\end{equation}
We define $T_p$ at which $I(T_p) = 0.34$ yielding $P \simeq 0.7$~\cite{Guth:1979bh,Guth:1981uk}.

\subsection{GW signals}

We now discuss the GW signals produced from the supercooled QCD -EW phase transitions induced from the topponium - Higgs correlation. 
We evaluate the two characteristic signal parameters, $\alpha$ and $\beta$. 
%The first one is the latent heat normalized to the radiation %energy density $\alpha$, which describes the strength of the %phase transition:
The normalized latent heat $\alpha$ is defined as 
\begin{equation}
    \alpha(T_p) \equiv \frac{1}{\rho_\text{rad}} \left( 
    \rho_{\rm rad}(T_i) + 
    \Delta V - T\frac{d(\Delta V)}{dT} \right)_{T = T_p}\,, 
\end{equation}
where $\rho_{\rm rad}(T_i) = \frac{\pi^2}{30} g_*(T_i) T_i^4$, which is removed depending on the assumed dark-sector supercooling scenario~\cite{Sagunski:2023ynd}. 
%we can approximately introduce the hidden-model-dependent $\alpha$
%\begin{equation}
%     \alpha(T_p) \simeq \frac{\rho_\text{rad}(T_i)}
%{\rho_\text{rad}(T_p)} = \left( \frac{T_i}{T_p} \right)^4.
%\end{equation}
The other parameter $\beta$ measures the inverse duration of the phase transition, and the normalized one ($\beta$) is defined as
\begin{equation}
    \tilde{\beta} \equiv \frac{\beta}{H(T_p)} = T_p \frac{d}{dT} \left( \frac{\mathcal{S}_3(T)}{T} \right).
\end{equation}

Given the signal parameters as above, 
we calculate the gravitational wave spectra sourced from the bubble collision. 
The power spectrum is given by~\cite{Brdar:2018num}
\begin{equation}
    \Omega_\text{coll} h^2 = e^{-4 N_e} \left( \frac{\rho_p}{\rho_r} \right) \times 1.67 \times 10^{-5} \left( \frac{\beta}{H(T_p)} \right)^{-2} \left( \frac{\kappa_\text{coll} \alpha}{1+\alpha} \right)^2 \left( \frac{100}{g_*} \right)^{1/3} \left( \frac{0.11 {v_w}^3}{0.42+{v_w}^2} \right) \frac{3.8(f/f_\text{coll})^{2.8}}{1+2.8(f/f_\text{coll})^{3.8}},
\end{equation}
where $\kappa_\text{coll}$ characterizes the energy transfer
between the vacuum energy and the kinetic energy of the bubble wall\cite{Zhang:2024vpp}, which we take $\kappa_\text{coll} \simeq 1$ with the velocity $v_\omega \simeq 1$;  $N_e$ denotes the e-folding number during the reheating and $\rho_{p,r}$ denotes the energy density at the percolation/reheating temperature. The peak frequency is given by~\cite{Zhang:2024vpp} 
\begin{equation}
    f_\text{coll} = e^{- N_e} \left( \frac{\rho_p}{\rho_r} \right)^{1/2} \times 1.65 \times 10^{-5} \left( \frac{\beta}{H(T_p)} \right) \left( \frac{0.62}{1.8-0.1v_w+{v_w}^2} \right) \left( \frac{T_r}{100 \, \text{GeV}} \right) \left( \frac{g_*}{100} \right)^{1/6}  \text{Hz}.
\end{equation}
In the present study, we simply assume the instantaneous reheating, such that 
in terms of the literature~\cite{Zhang:2024vpp}, 
$\rho_p= \rho_r$ and $N_e=0$. 

\begin{figure}[t]
    \centering
    \includegraphics[width=0.45\textwidth]{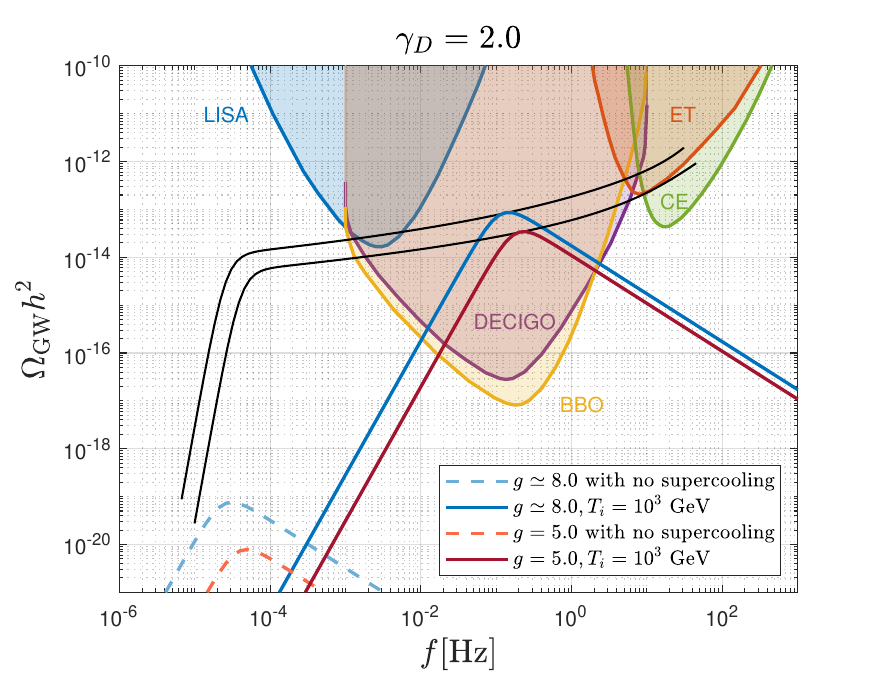}
     \includegraphics[width=0.45\textwidth]{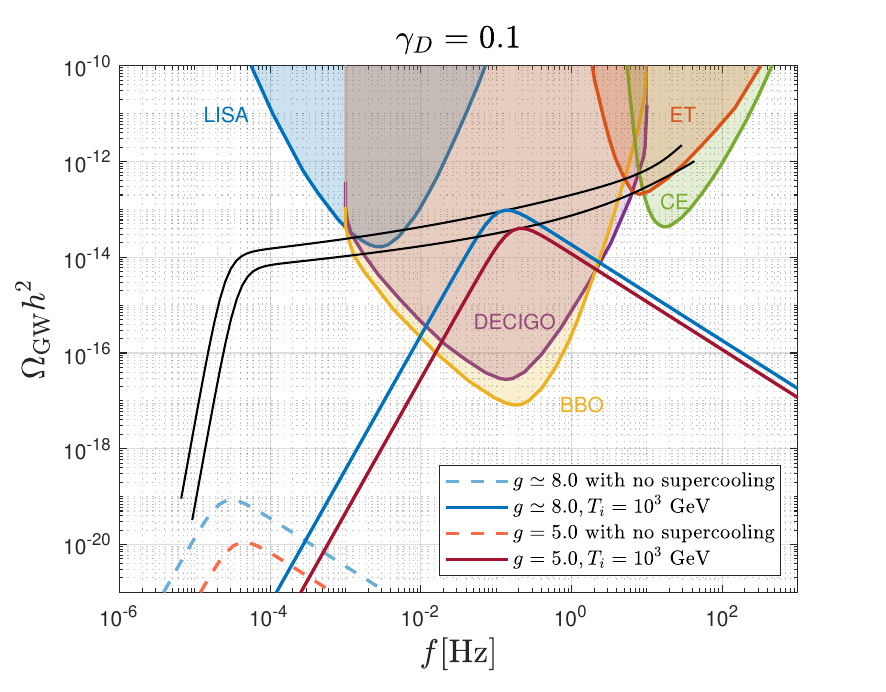}
    \caption{GW power spectra 
    compared to the sensitivities of the prospected GW interferometers~\cite{Schmitz:2020syl}. The black solid curves seen around the centers of the plots draw the trajectories following the shift of the peak frequency generated by varying $T_i$ from $1.5 \times 10^{-2}$ MeV to $1.0 \times 10^{6}$ GeV. The dashed curves around the bottom of the figure correspond to the case where the supercooling effect in the dark sector (to supply the EW scale) is no longer left in the QCD phase transition epoch.} 
    \label{fig:GW-signals}
\end{figure}

In Fig.~\ref{fig:GW-signals} we show the GW spectra predicted from 
the supercooling phase transition in the topponium - Higgs coupled system around 
the QCD phase transition epoch. 
As seen from the figure, we have observed no significant dependence 
of $\gamma_D$ in the range of $0 < \gamma_D \le 2$. 
The produced GW spectra can be probed by the LISA, DECIGO, BBO, CE, and ET depending 
on the dark-sector energy-density scale characterized by $T_i$. 
The signals will be present in addition to those arising from 
the supercooling in the dark sector (scalegenesis) phase transition as 
discussed in the literature~\cite{Iso:2017uuu,Hambye:2018qjv,Sagunski:2023ynd,Dichtl:2023xqd,Frandsen:2022klh,Ellis:2022lft,Bodeker:2021mcj,vonHarling:2017yew}. 
Just as a reference, also has been drawn the curves without the dark sector contribution (dashed curves around the bottom of the figure).  

%The reason why the solid line does not cross the peak point of the dashed lines is %the condition $T_i \gg T_p$ is no longer available, so the naive scaling would be a %little bit invalid.

\section{Summary and discussions}

In summary, we have made the first attempt to employ the nonperturbative analysis of the QCD-induced EW phase transition scenario, based on the LSD equations and the CJT formalism in the gauge-Higgs-Yukawa system. 
We have observed that the chiral broken QCD vacuum emerges with the nonperturbative top quark condensate, which can essentially be generated by the QCD sector: 
the EW phase transition at the QCD scale can be mainly triggered by QCD, even taking into account the strong top Yukawa interaction. 
Thus all the six quarks get masses small enough compared to 
the typical QCD critical temperature of the order of 100 MeV, where 
even the top quark can merely be in the same order (Fig.(\ref{fig:current-masses})). 
This is due to the big suppression of the top-Yukawa contributions to the ladder kernel caused by an accidental U(1) axial symmetry 
present in the top-Higgs Yukawa sector (Fig.\ref{fig:AvsP}).

We then have discussed the impact on the top quark condensation in the QCD-EW phase transition by employing a quark-meson model-like description as the low-energy 
effective theory. 
There the SM Higgs couples with the mean field of the top quark condensate and topponium-like scalar bound state. 
The nonperturbative scale anomaly induced from the dynamical top quark condensation based on the LSD analysis as well as the associated nonperturbative results have been encoded in the quark meson model, which highly constrains the model parameter space. 
we have observed the EW phase transition of the first order type along the 
topponium - Higgs hybrid direction in the field space (Figs.~\ref{fig:T-sigma}) and (\ref{fig:Omega}). 
We have discussed the associated GW productions and found that in addition to 
the conventional GW signals sourced from an expected BSM at around or over the TeV scale, the dynamical topponium-Higgs system can yield another power spectrum sensitive to the BBO and LISA, and DECIGO, etc (Fig.~\ref{fig:GW-signals}). 

Thus the presence of the nonperturbative top quark condensation in the supercooled EW phase transition would leave cosmologically probable footprints in the thermal history of the universe as the prospected GW spectra. 
This finding has presently been bench-marked based on the LSD method coupled with the quark meson-like and mean-field approach, which will pave a way to pursuing the QCD-origin cosmology coupled with possible BSM embedded in the classically scale-invariant scenario. 
In closing, we give several comments related to improvements from 
the present analysis, which is in more detail to be addressed in another publication.

\begin{figure}[t]
    \centering
    \includegraphics{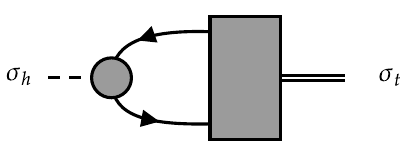}
\caption{A Feynman diagram generating the kinetic term mixing between $\sigma_t$ and 
$\sigma_h$ beyond the ladder approximation. The round blob attached on the $\sigma_h$ - $t$ - $t$ vertex, includes non-ladder corrections, in contrast to the ladder Fig.~\ref{fig:t-h-mix}, generating the $\langle \sigma_t \rangle$ - $\langle \sigma_h\rangle$ mixing, constructed only from the ladder graphs. }
\label{fig:nonL}
\end{figure}

\begin{itemize}

\item 

The present conclusion on the dominance of QCD to trigger the top condensation has been 
tied with the accidental $U(1)_A$ symmetry defined in the massless scalar theory, 
which is still good enough even in the presence of the scalar ($\sigma$) and pseudoscalar ($\pi_3$) 
mass difference, 
as has been noted in Sec.III.  
This symmetry is violated by a finite mass difference between $\sigma$ and $\pi_3$, which is actually 
generated even at the tree-level: $|m_\sigma^2 - m_{\pi_3}^2| = 2 \lambda_\phi \sigma^2_h$ (see also comments below 
Eq.(\ref{rota})). Thus the $U(1)_A$ symmetry is spontaneously broken (as in the same way as the $U(1)_Y$ in the ordinary SM) with nonzero Higgs VEV $\sigma_h$.  
As far as the perturbation theory with respect to the $\sigma$ and $\pi_3$ exchanges is concerned, 
however, the Yukawa interactions with $\sigma$ and $\pi_3$ get the same quantum-divergent corrections;  
$y_\sigma = y_\pi$ (up to finite parts). This is also related to the perturbative renormalizability 
of the spontaneously broken Higgs-Yukawa theory. 
Nonperturbative quantum corrections might separate the so-called effective charge  
$y_\sigma$ and $y_\pi$ for each fermion scattering amplitude into two different charges, $y_\sigma \neq y_\pi$. 
This would be noteworthy to do elsewhere. 
Note that this symmetry-protection argument has nothing to do with whether or not the ladder approximation is used.

\item 
In the present work, the thermal corrections to the top-Higgs system have been taken into account in the quark-meson model, where a couple of the results from the LSD analysis have been used as inputs to fix the model parameters. The straightforward 
incorporation of the thermal corrections in the LSD analysis is possible and provides 
the temperature dependence of the top and bottom quarks along with the Higgs VEV $\sigma_h$. Thus we could have discussed both QCD and EW phase transitions solely in the LSD framework, not invoking the mean field (or bosonization of $\bar{t}t$) $\sigma_t$ or passing the quark meson model. 
However, the cosmological phase transition issue currently seems to be challenging to argue without the bounce scalar action or the false vacuum decay via the scalar field theory, namely, without $\sigma_t$. This issue will be addressed in future work.

\item 
The LSD analysis can be improved by coupling to the functional renormalization group method, as has been discussed in~\cite{Gao:2020qsj}. 
It would be noteworthy to examine how non-ladder diagram corrections, incorporated into by the functional renormalization group, can affect the criticality of the top quark condensation and the thermal QCD and EW phase transitions.

\item 
The significant topponium effect on the EW phase transition arises from 
the mixing with the SM Higgs as depicted in Fig.~\ref{fig:t-h-mix}. 
This is essentially generated via the condensate mixture; $\langle \sigma_t \rangle$ 
- $\langle \sigma_h \rangle$, at the ladder approximation level, which thus contributes to the mean field potential as in Eq.(\ref{L:QMM}). 
The kinetic term mixing can also generically be generated in the full gauge-Higgs-Yukawa theory, however, which goes beyond the ladder approximation because 
it is necessary to include the Yukawa vertex corrections. See Fig.~\ref{fig:nonL}. 
This beyond-the-ladder kinetic term mixing at finite temperature 
would modify the bounce action Eq.(\ref{BA}), which might give a nontrivial 
correction to the nucleation and percolation processes in the cosmological 
phase transition. Similar kinetic term corrections have been addressed 
in the literature~\cite{Sagunski:2023ynd,Aoki:2019mlt} based on the Nambu-Jona-Lasinio type model for a dark QCD theory, in which, however, the kinetic mixing effect has been disregarded in the bounce equation.  
This issue is intriguing to be discussed in detail elsewhere.

\end{itemize}

\section*{Acknowledgments} 

We are grateful to Fei Gao, and He-Xu Zhang for fruitful discussions. 
We would like to present a special thanks to Masatoshi Yamada for enlightening discussions and comments. 
This work was supported in part by the National Science Foundation of China (NSFC) under Grant No.11747308, 11975108, 12047569, and the Seeds Funding of Jilin University.

\bibliographystyle{elsarticle-num}
\bibliography{references}

\end{document}